\documentclass{elsart}

\usepackage{amssymb}
\usepackage{amsmath}

\newcommand{\bra}[1]{\langle #1|}
\newcommand{\ket}[1]{|#1\rangle}
\newcommand{\braket}[2]{\langle #1|#2\rangle}

\def\Nj{\{N_j\}}

\def\d{{\rm d}}
\def\i{i}
\def\parti{n} 
\def\Pro{{\sf P}}

\newcommand{\tr}{{\rm tr}}

\def\xiv{\mbox{\boldmath$\xi$}}

\newcommand{\p}{{\rm p}}
\newcommand{\x}{{\rm x}}

\begin{document}

\begin{frontmatter}



\title{The microcanonical ensemble of the ideal relativistic quantum gas}


\author{F. Becattini, L. Ferroni}

\address{Universit\`a di Firenze and INFN Sezione di Firenze}

\begin{abstract}
We derive the microcanonical partition function of the ideal relativistic 
quantum gas of spinless bosons in a quantum field framework as an expansion 
over fixed multiplicities. Our calculation generalizes well known expressions 
in literature in that it does not introduce any large volume approximation 
and it is valid at any volume. We discuss the issues concerned with the 
definition of the microcanonical ensemble for a free quantum field at volumes 
comparable with the Compton wavelength and provide a consistent prescription 
of calculating the microcanonical partition function, which is finite at finite 
volume and yielding the correct thermodynamic limit. Besides an immaterial 
overall factor, the obtained expression turns out to be the same as in the 
non-relativistic multi-particle approach. This work is introductory to derive 
the most general expression of the microcanonical partition function fixing 
the maximal set of observables of the Poincar\'e group.
\end{abstract}

\end{frontmatter}

\section{Introduction}

The microcanonical ensemble of the relativistic gas is a subject which has not 
received much attention in the past. The reason of scarce interest in this problem 
is the peculiarity of physical applications, which have been essentially confined 
within statistical approaches to hadron production and the bag model \cite{bag};
these are indeed the only cases where involved volumes and particle numbers
are so small that microcanonical corrections to average quantities become 
relevant. Otherwise, the involved energies or volumes are so large that canonical and 
grand-canonical ensembles are appropriate for most practical purposes (e.g. in
relativistic heavy ion collisions).
Recently \cite{goren}, it has been pointed out that the equivalence, in the thermodynamic
limit, between grand-canonical, canonical and microcanonical ensembles does not 
apply to fluctuations, more in general to moments of multiplicity distributions of
order $ > 1$. Indeed, effects of the difference between statistical ensembles might 
be unveiled in studying multiplicity distributions in relativistic nuclear collisions,
\cite{begun}. In view of these phenomenological applications, it would be then desirable
to have an in-depth analysis of the microcanonical ensemble of a relativistic
quantum gas.

The main difficulty in tackling this problem stems from the need of imposing a finite 
volume. This is necessary to have a correct thermodynamic limit because, being the 
energy $E$ finite by construction, also $V$ must be finite if the limit with $E/V$ 
fixed is to be taken. Strange as it may seem, a full and rigorous treatment of the 
relativistic microcanonical ensemble of an ideal gas at finite volume is still 
missing. In all previous works on the subject, at some point, the large volume 
approximation is introduced; this is tacitly done, for instance, considering the 
single-particle level density as continuous, namely replacing sums over discrete 
quantum states with momentum space integrations \cite{chh}:
\begin{equation}
  \sum_{\bf k} \rightarrow \int \d^3 \p \; .
\end{equation} 
In a previous work \cite{bf1}, we have derived an expression of the microcanonical 
partition function of an ideal relativistic quantum gas with explicit finite-volume 
corrections (see eqs.~(\ref{singlespecies}),(\ref{multispecies})). However, that 
expression was obtained in an essentially 
multi-particle first-quantization framework, which, as pointed out in ref.~\cite{meaning} 
should be expected to become inadequate at very low volumes, comparable with the 
Compton wavelength of particles. In this regime, underlying quantum field effects 
should become important and pair creation due to localization an unavoidable effect 
\cite{landau}. 
Indeed, there are several studies of the microcanonical ensemble of a free quantum 
field \cite{prevqft}, but all of them, again, at some point invoke a large-volume 
approximation. In the limit of large volumes one obtains the same expressions of 
the microcanonical partition function and, consequently, of statistical averages 
as in the first-quantization multi-particle approach followed in ref.~\cite{bf1}.

The aim of this work is to derive a general expression of the microcanonical
partition function in a full relativistic quantum field framework, valid for {\em any
finite volume}, generalizing the results obtained in~\cite{bf1}. We will do this 
for the simplest case of an ideal gas of spinless bosons and postpone the treatment of
particles with spin to a forthcoming publication \cite{bfprep}. We will show that 
the expression of the microcanonical partition function obtained in ref.~\cite{bf1} 
in a non-relativistic multi-particle approach holds provided that a consistent 
prescription of subtracting terms arising from field degrees of freedom outside 
the considered volume is introduced. 

The paper is organized as follows: in Sects.~2 and 3 we will argue about general 
features of the microcanonical ensemble for a relativistic system and discuss several 
issues concerning a proper definition of the microcanonical partition function. 
In Sect.~4 we will cope with the further issues related to the definition of a 
microcanonical ensemble for a quantum field at finite volume. Sects.~5 and 6 
include the main body of this work, where the microcanonical partition function
is worked out in a quantum field theoretical framework. In Sect.~7 we will summarize
and discuss the results.

\section{On the definition of the microcanonical partition function}

It is well known that the fundamental tool to calculate statistical averages in any
ensemble is the partition function. For the microcanonical ensemble one has to calculate
the {\em microcanonical partition function} (MPF) which is usually defined as the 
number of states with a definite value $E$ of total energy: 
\begin{equation}
 \Omega \equiv \sum_{{\rm states}} \delta(E-E_{{\rm state}}) \; .
\end{equation}
For a quantum system, the MPF is the trace of the operator $\delta(E-\hat H)$:
\footnote{Throughout this work quantum operators will be distinguished from 
ordinary numbers by a symbol `` $\hat{}$ ''.}
\begin{equation}\label{mpf0}
 \Omega = \tr \delta(E - \hat H)
\end{equation}
with proper normalization of the basis states. For instance, for one non-relativistic
free particle, one has to calculate the trace summing over plane waves normalized 
so as to $\braket{{\bf p}}{{\bf p}'} = \delta^3({\bf p}-{\bf p}')$:
\begin{equation}\label{1part}
 \Omega = \tr (E-\hat H) = \sum_{\bf p} \delta \left(E - \frac{\p^2}{2m}\right) 
  \braket{{\bf p}}{{\bf p}} =
  \frac{1}{(2\pi)^3}\int \d^3 \x \int \d^3 \p \, \delta \left(E - \frac{\p^2}{2m}\right)
\end{equation}
Thereby, one recovers the well known classical expression implying that the MPF
is the number of phase space cells with size $h^3$ and given energy $E$. This
number is infinite as the volume is unbounded and it is thus impossible
to calculate a meaningful thermodynamic limit at finite energy density. Hence, 
one usually considers a system confined within a finite region by modifying the 
hamiltonian $\hat H$ with the addition of infinite potential walls, i.e. setting 
$\hat H' = \hat H + \hat V$ where $\hat H$ is the actual internal hamiltonian 
and $\hat V$ an external potential implementing infinite walls. Classically, 
this leads to a finite $\Omega$, namely:
\begin{equation}\label{classic}
 \Omega = \frac{V}{(2\pi)^3} \int \d^3 \p \, \delta \left(E - \frac{\p^2}{2m}\right)
\end{equation}
where $V$ is the volume of the region encompassed by the potential 
walls.\footnote{We will use the same symbol $V$ to denote both the finite region and 
its volume.}
Also the corresponding quantum problem can be easily solved and one has:
\begin{equation}\label{discrete}
 \Omega = \tr (E-\hat H') \equiv \sum_{\bf k} \delta 
 \left(E - \frac{{\rm k}^2}{2m}\right) 
\end{equation}
where the sum runs over all wave vectors ${\bf k}$ which, for a rectangular box
with side $L_i$ and periodic boundary conditions, are labelled by three integers 
$(n_1,n_2,n_3)$ such that ${\rm k}_i = n_i \pi/L_i$. The difficulty of the quantum
expression (\ref{discrete}) with respect to the classical one (\ref{classic}) is 
that, for a given energy $E$, a set of integers fulfilling the constraint imposed 
by the Dirac's $\delta$:
$$
 \frac{\pi^2}{2m}\left({n^2_x \over L^2_x}+{n^2_y \over L^2_y}+{n^2_z \over L^2_z}
\right) = E
$$
in general does not exist. Therefore, the MPF vanishes except for a discrete set 
of total energies, for which it is divergent. One has a finite result only for 
the integral number of states
$\int_0^{E'} \d E \, \Omega(E)$, that is the number of states with an energy less 
than a given $E'$, but this is clearly a stepwise and non-differentiable function of 
$E'$. 

This holds for an ideal gas of any finite number of particles: strictly speaking, 
the MPF cannot be defined at finite
energy and volume as a continous function. Only in the thermodynamic limit $E\to \infty$
and $V \to \infty$ an expression like (\ref{discrete}) becomes meaningful because
it is then possible to replace the sum over discrete levels with a phase space 
integration: 
\begin{equation}\label{sumocell}
  \sum_{{\rm cells}} \underset{V \to \infty}{\longrightarrow}
  \frac{V}{(2\pi)^3} \int \d^3 \p \; .
\end{equation}  
Therefore, for a truly finite quantum system, one needs a better definition 
of microcanonical partition function. A definition which does not suffer from 
previous drawbacks is the following: 
\begin{equation}\label{mpf1}
 \Omega =  \tr_V \delta(E-\hat H) \equiv \sum_{h_V} \bra{h_V} \delta(E-\hat{H}) 
\ket{h_V} \; . 
\end{equation}
where $\hat H$ is the internal hamiltonian, without external confining potential,
and the $\ket{h_V}$'s form a complete set of normalized {\em localized states}, 
i.e. a complete set of states for the wavefunctions vanishing out of the region 
$V$. It should be stressed that these states are not a basis of the full Hilbert 
space because wavefunctions which do not vanish out of $V$ cannot be expanded 
onto this basis; thence the  
notation $\tr_V$ instead of $\tr$ meaning that the trace in eq.~(\ref{mpf1}) is 
not a proper one. The difference between (\ref{mpf1}) and a definition like 
(\ref{mpf0}) 
is that $\ket{h_V}$ are not eigenstates of the hamiltonian $\hat H$ and the right hand 
side of (\ref{mpf1}) does not reduce to a discrete sum of $\delta$'s. In fact, 
this is crucial to have a continuous function of $E$, unlike (\ref{discrete}). 

As an example, let us work out the definition (\ref{mpf1}) for the single free 
particle confined in a rectangular box by infinite potential walls and compare 
it to (\ref{1part}). A complete set of states for this problem is:
\begin{equation}\label{complset}
  \ket{{\bf k}} = \begin{cases} 
                  \frac{1}{\sqrt V}\exp(\i {\bf k} \cdot {\bf x}) \qquad \; 
                  \text{if ${\bf x}\in V$}
                  \qquad{\bf k}= \pi n_x/L_x \pi {\bf \hat i} +
                   n_y/L_y {\bf \hat j} \pi n_z/L_z {\bf \hat k} \\
                  0 \qquad \qquad \qquad \qquad \text{if ${\bf x} \notin V$} 
                  \end{cases}
\end{equation}
Therefore, the MPF definition (\ref{mpf1}) implies:
\begin{eqnarray}\label{examp1}
  \Omega = \sum_{\bf k} \bra{{\bf k}} \delta(E-\hat H) \ket{{\bf k}} =
   \sum_{\bf k} \int \d^3 \p \, |\braket{{\bf k}}{{\bf p}}|^2 
   \delta\left(E - \frac{\p^2}{2m}\right)  
\end{eqnarray}
where we have inserted a resolution of the identity by using a complete set of 
states for the full Hilbert space. The sum in (\ref{examp1}) can be calculated and
yields:
\begin{eqnarray}\label{examp2}
&&\sum_{{\bf k}} |\braket{{\bf k}}{{\bf p}}|^2 = \sum_{{\bf k}} \frac{1}{V (2\pi)^3} 
 \Big| \int_V \d^3 \x \, \exp[\i({\bf k - p})\cdot {\bf x}] \Big|^2  \nonumber \\
&& = \sum_{{\bf k}} \frac{1}{V} \int_V \d^3 \x \int_V \d^3 \x' \, 
  \exp[\i{\bf k}\cdot ({\bf x}-{\bf x}')] \exp[-\i{\bf p}\cdot ({\bf x}-{\bf x}')] 
 \nonumber \\
&& =  \frac{1}{(2\pi)^3} \int_V \d^3 \x \int_V \d^3 \x' \, \delta^3({\bf x}-{\bf x}')] 
  \exp[-\i{\bf p}\cdot ({\bf x}-{\bf x}')] = \frac{V}{(2\pi)^3}
\end{eqnarray}
where the completeness relation in $V$:
\begin{equation}
 \sum_{\bf k} \frac{1}{V} \exp[\i {\bf k}\cdot ({\bf x}-{\bf x}')] = 
  \delta^3({\bf x}-{\bf x}') 
\end{equation}
has been used. Thus, by using (\ref{examp2}), the eq.~(\ref{examp1}) turns into:
\begin{equation}\label{examp3}
  \Omega = \frac{V}{(2\pi)^3} \int \d^3 \p \, \delta\left(E - \frac{\p^2}{2m}\right)  
\end{equation}
that is the same expression (\ref{classic}) as in the classical case. 

The MPF (\ref{examp3}) is now manifestly a continuous function of $E$ and, remarkably, 
its thermodynamic limit $V \to \infty$ is the same as the thermodynamic limit of the 
``pure" quantum expression (\ref{discrete}) (because of (\ref{sumocell})). Since 
the only strict requirement for a well defined MPF is to reproduce the correct
thermodynamic limit, for a gas one can choose a definition like (\ref{mpf1}) 
instead of $\Omega = \tr \delta (E - \hat H')$ in (\ref{discrete}). We emphasize 
again that in the passage from (\ref{discrete}) to (\ref{mpf1}) the hamiltonian 
embodying an external confining potential is replaced with the internal hamiltonian 
while, at the same time, the localized eigenstates of the former hamiltonian 
are used to calculate the trace. 

\section{The microcanonical partition function of a relativistic system}

In special relativity, the microcanonical ensemble must include momentum conservation 
beside energy's to fulfill Lorentz invariance. This means that the MPF definition 
(\ref{mpf1}) should be generalized to \cite{bf1}:
\begin{equation}\label{micro1}
 \Omega= \sum_{h_V}  \bra{h_V} \delta^4(P-\hat{P}) \ket{h_V} \; , 
\end{equation}
$P$ being the four-momentum of the system and $\hat{P}$ the four-momentum operator. 
The MPF now being a number of states per four-momentum cell, it is a Lorentz-invariant 
quantity. The calculation of the MPF is easiest in the rest-frame of the system,
where $P=(M,{\bf 0})$ and the four-volume $V u$, $u$ being the four-velocity and $V$ 
the proper volume of the system, reduces to (V,{\bf 0}), according to the usual 
formulation of statistical relativistic thermodynamics \cite{touschek,karsch}. 

The eq.~(\ref{micro1}) can be further generalized by enforcing the conservation
of not only energy-momentum but of the maximal set of conserved quantities, i.e. 
a maximal set of commuting observables built with the generators of the Poincar\'e 
group. To achieve this, one has to replace $\delta^4(P-\hat{P})$ in (\ref{micro1})
with a generic projector $\Pro_i$ over an irreducible state of the representation
of the Poincar\'e group \cite{bf1,meaning}, i.e:  
\begin{equation}\label{mpf}
 \Omega = \sum_{h_V} \bra{h_V} \Pro_i \ket{h_V} \; .
\end{equation}
This ensemble is still generally defined as microcanonical ensemble and (\ref{mpf})
microcanonical partition function. 

In this work, we will confine to a microcanonical ensemble where only energy-momentum
is fixed, i.e. our projector ${\sf P}_i$ in eq.~(\ref{mpf}) will be:
\begin{equation}\label{t4}
 \Pro_i = \delta^4(P - \hat P) 
\end{equation}
and to an ideal quantum gas, i.e. with $\hat P$ being the free four-momentum operator.
In fact, it should be stressed that $\delta^4(P -\hat{P})$ is not a proper projector, 
because ${\sf P}^2 = a{\sf P}$ where $a$ is a divergent constant. This is owing
to the fact that normalized projectors onto irreducible representations cannot be 
defined for non-compact groups, such as the space-time translation group T(4). 
Nevertheless, we will maintain this naming even for non-idempotent operators, 
relaxing mathematical rigor, because it will be favorable to adopt the projection 
formalism. 

It is worth pointing out that the definition (\ref{mpf1}) involving only the internal
(free) hamiltonian, is much more fit than (\ref{discrete}) for a relativistic generalization. 
Besides the advantage of restoring continuity in $E$, discussed in previous 
section, this formulation can be easily extended to the full set of 
conservation laws without major conceptual difficulty. Conversely, had one tried to 
generalize (\ref{discrete}), one should have defined a finite region and {\em afterwards} 
sought the observables commuting with the hamiltonian supplemented with an external 
confining potential. This would have not been an easy task, and, moreover, a maximal 
set of observables commuting with the modified $\hat H$ would not, in general, 
define a Poincar\'e algebra. This is a well known problem in the static bag model 
where the translational invariance is manifestly broken and momentum is thus not 
conserved. On the other hand, in the definition 
(\ref{mpf}), we deal with the original Poincar\'e algebra of unmodified (free) 
operators and enforce the localization through the projector onto confined states. 

The eq.~(\ref{mpf}) can be recast as a full trace by inserting a complete set of 
states $\ket{f}$ into (\ref{mpf}): 
\begin{eqnarray}\label{mpf2}
 \Omega &=& \sum_{h_V} \bra{h_V} \sum_f \ket{f} \bra{f} \Pro_i \ket{h_V} \nonumber \\
        &=& \sum_f \bra{f} \Pro_i \sum_{h_V} \ket{h_V} \braket{h_V}{f} \equiv 
	\sum_f \bra{f} \Pro_i \Pro_V \ket{f} = \tr \left[ \Pro_i \Pro_V \right]
\end{eqnarray}
where 
\begin{equation}\label{prov}
 \Pro_V = \sum_{h_V} \ket{h_V}\bra{h_V} \; ,
\end{equation}
is, by definition, the projector onto the Hilbert subspace $H_V$ of confined 
states (i.e. of wavefunctions vanishing out of $V$). 
The formula (\ref{mpf2}) is the starting point to carry out a calculation of the 
MPF at finite volume. The first thing to do is to expand (\ref{mpf2}) as a sum of 
partition functions at fixed multiplicities, i.e.:
\begin{equation} \label{multexp}
\Omega= \sum_{N} \Omega_N 
\end{equation}
for a single species gas and:
\begin{equation}\label{sumovchann}
 \Omega = \sum_{\Nj} \Omega_{\Nj}
\end{equation}
for a multi-species gas, where $\Nj=(N_1,\ldots,N_K)$ is a set of particle 
multiplicities for each species $j = 1,\ldots,K$, defining a {\em channel}. 
$\Omega_N$ or $\Omega_{\Nj}$ are obtained by summing over all possible values of 
kinematical variables with fixed multiplicities. So, if $\ket{f} \equiv 
\ket{N,\{ p \}}$ where $\{ p \}$ labels the set of kinematical 
variables of particles in the state $\ket{f}$, $\Omega_N$ reads: 
\begin{equation} \label{multexp2}
\Omega_N= \sum_{\{ p \} } \bra{ N,\{ p \}} \Pro_i \Pro_V  \ket{ N,\{ p \}} \; . 
\end{equation}
Likewise, for a multi-species gas, the microcanonical partition function is expressed 
as an expansion over all possible channels:
\begin{equation}\label{mpf2summ}
 \Omega_{\Nj} = \sum_{\{ p \}} \bra{ \Nj,\{ p \}} \Pro_i \Pro_V  \ket{\Nj,\{ p \}} \; .
\end{equation}
and $\Omega_{\Nj}$ is defined as the {\em microcanonical channel weight}.

The microcanonical channel weights (\ref{mpf2summ}) have been calculated in 
ref.~\cite{bf1} with energy-momentum conservation (i.e. using (\ref{t4}) in a 
multi-particle, first quantization framework. For a single species ideal spinless
gas:
\begin{equation}\label{singlespecies}
\Omega_{N} = \frac{1}{N!} \int \d^3 \p_1 \ldots \d^3 \p_N \,
 \delta^4 \left( P - \sum_{\parti=1}^{N} p_{\parti} \right) 
 \sum_{\rho \in \; {\rm S}_{N}} \prod_{\parti=1}^{N} 
  F_V({\bf p}_{\rho({\parti})}-{\bf p}_{\parti}) 
\end{equation}
while for a multi-species gas of $K$ spinless bosons:
\begin{eqnarray}\label{multispecies}
\Omega_{\Nj} &=& \left[\prod_{j=1}^{K}\frac{1}{N_j!} \prod_{\parti_j=1}^{N_j} 
 \int \d^3 \p_{\parti_j}  \right] 
 \delta^4 \left( P - \sum_{\parti=1}^{N} p_{\parti} \right) \nonumber \\
 &\times& \prod_{j=1}^{k}\sum_{\rho_j \in \; {\rm S}_{N_j}}\prod_{\parti_j=1}^{N_j} 
 F_V({\bf p}_{\rho_j({\parti_j})}-{\bf p}_{\parti_j}) 
\end{eqnarray} 
being $N=\sum_j N_j$.
In eqs.~(\ref{singlespecies}),(\ref{multispecies}) $\rho_j$ labels permutation 
belonging to the permutation group ${\rm S}_{N_j}$ and $F_V$ are Fourier integrals 
over the region $V$:
\begin{equation}\label{fourier2}
  F_V({\bf p} - {\bf p}') \equiv 
 \frac{1}{(2\pi)^3} \int_V \d^3 \x \; \e^{ i ({\bf p} - {\bf p}')\cdot {\bf x}} 
\end{equation}
If the volume is large enough so as to allow the approximation:
\begin{equation}\label{delta}
  F_V({\bf p} - {\bf p}') = 
   \frac{1}{(2\pi)^3} \int_V \d^3 \x \; \e^{ i ({\bf p} - {\bf p}')\cdot 
  {\bf x}} \simeq \delta^3 ({\bf p} - {\bf p}')
\end{equation}
the microcanonical channel weights (\ref{multispecies}) can be resummed explicitely 
into the microcanonical partition function according to (\ref{sumovchann}) and one 
obtains \cite{bf1}:
\begin{equation}\label{omold}
\Omega = \frac{\lim_{\varepsilon \to 0}}{(2 \pi)^4} 
\int_{-\infty-\i \varepsilon}^{+\infty-\i \varepsilon} \!\!\!\!\!\!\!\!\! 
 \d y^0 \int \d^3 {\rm y} \; \e^{\i P \cdot y} \exp \left[ \sum_j \frac{V}
 {(2 \pi)^3} \int \d^3 \p \; \log (1 - \e^{-\i p \cdot y})^{-1} \right] \; .
\end{equation}
A full analytical calculation of 
eq.~(\ref{omold}) is possible only for the limiting case of vanishing masses (e.g. 
microcanonical black body). For the massive case, four-dimensional integrations cannot 
be worked out analytically and one has to resort to numerical computation.

The eq.~(\ref{omold}) was implicitely obtained in ref.~\cite{chh} where the first
expression of the microcanonical partition function of a multi-species ideal relativistic 
quantum gas was derived as an expansion (\ref{sumovchann}) over channels, by using 
the large-volume approximation (\ref{sumocell}) from the very beginning. This shows 
that the approximation (\ref{delta}) is indeed equivalent to the (\ref{sumocell}), 
as also demonstrated in ref.~\cite{bf1}. Noticeably, the MPF definition eq.~(\ref{micro1}) 
without any large volume approximation involves the appearance of Fourier integrals 
accounting for Bose-Einstein and Fermi-Dirac correlations in the quantum gas, which 
do not show up in the large-volume approximation enforced in ref.~\cite{chh}. This approach
also allows to investigate further generalizations when the volume is so small that 
relativistic quantum field effects must be taken into account. 
  
\section{Microcanonical ensemble and field theory}\label{sec:micfield}

The calculation of the partition function (\ref{micro1}) in a quantum field framework
brings in new difficulties with respect to the first-quantization scheme. This 
problem has been approached in literature with a functional approach, 
inspired of the usual grand-canonical thermal field theory \cite{prevqft}. 
However, these calculations aim at the limit of large volumes and are therefore
insensitive to the difficulties pertaining to the strict requirement of finite 
volume discussed in detail in Sect.~2. As a result, for a free field, the derived 
expressions are equivalent to the formula (\ref{omold}). 

Instead of starting with a functional integration from the very beginning, we 
calculate the microcanonical partition function of a free field by first expanding 
it at fixed multiplicities like in eqs.~(\ref{multexp}),(\ref{multexp2}) 
(or channels, for multi-species gas like in eq.~(\ref{sumovchann})), where 
$\ket{N,\{ p \}}$ are Fock space states with definite particle multiplicity
and kinematical variables $\{p\}$. To carry out this calculation, we first need to 
find an expression of the {\em microcanonical state weight}:
\begin{equation} \label{density}
\omega \equiv \bra{N,\{ p \}} \Pro_i \Pro_V \ket{N,\{ p \}} \; . 
\end{equation}
By using (\ref{t4}) and choosing $\ket{N,\{ p \}}$ as an eigenstate of the 
four-momentum operator with eigenvalue $P_f = \sum_i p_i$, the 
eq.~(\ref{density}) becomes:
\begin{equation}\label{t4density}
  \omega = \delta^4(P - P_f) \, \bra{N,\{ p \}} \Pro_V \ket{N,\{ p \}} \; ,
\end{equation}
To calculate $\omega$ and, by a further integration, $\Omega_N$ we need to know 
the projector $\Pro_V$. Since $\Pro_V$ is defined as the projector onto the 
Hilbert subspace of localized states, it can be easily written down in a 
multi-particle non-relativistic quantum mechanical (NRQM) framework. As an example, 
for a non-relativistic spinless single particle, it reads (see also Sect.~2):
\begin{equation}\label{prosingle}
\Pro_V = \sum_{k_V} \ket{k_V}\bra{k_V}
\end{equation}
where $\ket{k_V}$ is a normalized state of the particle confined in a region 
$V$, with a corresponding wavefunction $\psi_{k_V}({\bf x})$ vanishing out of $V$.
The symbol $k_V$ stands for a set of three numbers labelling the kinematical modes 
of the confined states (e.g. discrete wavevectors, or energy and angular momenta)
and the set $\ket{k_V}$ form a complete set of states for the wavefunctions 
vanishing out of $V$. The projector (\ref{prosingle}) can be easily 
extended to the many-body case and: 
\begin{equation}\label{pv1}
 \Pro_V = \sum_{\widetilde{N},\{ k \} } \ket{\widetilde{N},\{ k_V\}}
 \bra{\widetilde{N},\{ k_V \}} \; .
\end{equation}
where the symbol $\{ k_V \}$ denotes a multiple set of kinematical modes of the 
confined states while $\widetilde{N}$ is the integrated occupation number, 
i.e. the sum of occupation numbers over all single-particle kinematical modes. 
In the NRQM approach these numbers are simply particles multiplicities, implying:
\begin{equation}\label{condition}
  \braket{N,\{ p \}}{\widetilde{N},\{ k_V \}} \ne 0 \qquad {\rm iff} 
  \;\; N = \widetilde{N} \; .
\end{equation}
To calculate the microcanonical state weight, thence the MPF, the products 
$\braket{\widetilde{N},\{ k_V \}}{N,\{ p \}}$ can be worked out on the basis of 
(\ref{condition}) similarly to what has been done in Sect.~2 for a single particle,
yielding, for a scalar boson \cite{bf1}:
\begin{equation}\label{clust}
 \bra{N,\{ p \}} \Pro_V \ket{N,\{ p \}}= \sum_{\rho \in \,{\rm S}_{N}} 
 \prod_{{\parti}=1}^{N} F_V({\bf p}_{\rho(\parti)}-{\bf p}_{\parti}) 
\end{equation}
where $\rho$ is a permutation of the integers $1,\ldots,N$ and $F_V$ are Fourier 
integrals (\ref{fourier2}) over the system region $V$. From the above equation
the expression of the microcanonical channel weight in eq.~(\ref{singlespecies}) 
follows. We will refer to the expression (\ref{clust}) as the NRQM one, meaning 
that is has been obtained in this first-quantized multi-particle NRQM framework, 
where, for instance, particles and antiparticles are simply considered as different 
species and their contributions factorize.
  
One could envisage that a projector like (\ref{pv1}), written in terms of Fock 
space states, could be simply carried over to the relativistic quantum field case,
where $\ket{\widetilde{N},\{ k_V\}}$ are states of the localized problem, obtained
by solving the free field equations in a box with suitable boundary conditions.  
Yet, some difficulties soon arise. First of all, whilst in NRQM the single-particle 
localized wavefunction $\ket{k_V}$ and the free plane wave 
state $\ket{{\bf p}}$ live in the same Hilbert space, in quantum field theory the 
localized and the non-localized problem are associated with distinct Hilbert spaces. 
Thus, unlike in NRQM, it is not clear how to calculate
a product like $\braket{\widetilde{N},\{ k_V \}}{N,\{ p \}}$. Secondly, even if 
there was a definite prescription for it, it should be expected that {\em the 
integrated occupation numbers of the localized problem do not coincide with actual 
particle multiplicities unless the volume is infinite}. To understand this point, 
one should keep in mind that properly called particles arise from solutions of 
the free field equations over the whole space and that the hamiltonian eigenstates 
of the localized problem are conceptually different. Consequently, the integrated
particle number operators in the whole space should differ from integrated number 
operators within the finite region. Hence, unlike in NRQM, a state with definite 
integrated occupation numbers $\widetilde N$ (we purposely refrain from calling 
them particle numbers) should be expected to have non vanishing components on 
{\em all} free states with different numbers of actual particles, namely:
\begin{equation}\label{confstates}
\ket{\widetilde{N}}_V = \alpha_{0,\widetilde{N}}\ket{0}+
 \alpha_{1,\widetilde{N}}\ket{1}+\ldots+
 \alpha_{\widetilde{N},\widetilde{N}} \ket{\widetilde{N}}+\ldots   
\end{equation}
where $\alpha_{i,\widetilde{N}}$ are non-trivial complex coefficients, and 
(\ref{condition}) no longer holds. Only in the large volume limit one expects
that the integrated occupation numbers coincide with actual multiplicities and
eq.~(\ref{condition}) applies. This kind of effect is pointed out in the introduction 
of Landau's book on quantum field theory \cite{landau}: when trying to localize 
an electron, an electron-positron pair unavoidably appears, meaning that the 
localized single ``particle" is indeed a superposition of many true, asymptotic 
particle states. Another relevant manifestation of this difference 
which is probably more familiar, is the Casimir effect, which is related to the 
difference between the true vacuum state $\ket{0}$ and the localized vacuum state 
$\ket{0}_V$. Hence, all formulae derived under the approximation (\ref{condition}) 
are asymptotic ones, valid in the limit $V \to \infty$ but not at strictly finite 
volume. Thus, one should expect significant finite-volume corrections to the 
eqs.~(\ref{clust}) and the ensuing MPF (\ref{singlespecies}) in a quantum field
treatment.

If we want to give an expression like $\braket{\widetilde{N},\{ k_V \}}{N,\{ p \}}$
a precise meaning in a quantum field framework, we first need to map the Hilbert 
space $H_V$ of the localized problem into the Hilbert space $H$ of the free field 
over the whole space. This can be done by mapping the field eigenstates and operators 
of $H$ into $H_V$ in a natural way as:
\begin{eqnarray}\label{mapping}
&&  \Psi({\bf x})_{H_V} \longrightarrow \Psi({\bf x})_{H} \nonumber \\
&&  \ket{\psi({\bf x})}_{H_V} \longrightarrow \ket{\psi({\bf x})}_{H} 
\end{eqnarray}
This allows writing linear, non-bijective, Bogoliubov relations expressing the
annihilation and creation operators of the finite region problem as a function of
those of the whole (real scalar) field (see Appendix A for derivation):
\begin{equation}\label{bogo}
 a_{{\bf k}} = \int \d^3 \p \, F({\bf k},{\bf p}) 
 \frac{\varepsilon_{{\bf k}} + \varepsilon_{{\bf p}}}{2\sqrt{\varepsilon_{{\bf k}}\varepsilon_{{\bf p}}}}
 \, a_{{\bf p}} + F({\bf k},-{\bf p}) 
 \frac{\varepsilon_{{\bf k}} - \varepsilon_{{\bf p}}}{2\sqrt{\varepsilon_{{\bf k}}\varepsilon_{{\bf p}}}}
 \, a_{{\bf p}}^\dagger
\end{equation}
where $\bf k$ are triplets of numbers labelling kinematical modes, just like the
aforementioned $k_V$, $\varepsilon_{{\bf k}}$ is the associated energy, 
$\varepsilon_{{\bf p}} = \sqrt{\p^2+m^2}$ and:
\begin{equation}
 F({\bf k},{\bf p}) = \frac{1}{(2\pi)^3} 
  \int_V \d^3 \x \, u_{\bf k}^*({\bf x}) \e^{i {\bf p}\cdot{\bf x}}   
\end{equation}
$u_{{\bf k}}$ being a complete set of orthonormal wavefunctions for the finite region. 
A remarkable feature of relativistic quantum fields is that, unlike in NRQM, the 
localized annihilation operators have non-vanishing components onto the creation 
operators in the whole space, as shown by (\ref{bogo}). This 
confirms our expectation that a localized state with definite integrated occupation 
numbers is a non-trivial linear combination of states with different particle 
multiplicities. Expectedly, as the volume increases, these components become 
smaller and in the infinite volume limit one recovers $a_{{\bf k}} = a_{{\bf p}}$
(see Appendix A).

Starting from the Bogoliubov relations (\ref{bogo}), it should be possible, in 
principle, to calculate the coefficients in eq.~(\ref{confstates}), thence the 
microcanonical state weight (\ref{t4density}) by using the expansion (\ref{pv1}). 
In fact, we do not really need to do that. It is more advantageous, as pointed out in 
ref.~\cite{meaning}, to write the projector $\Pro_V$ in terms of field states 
rather than occupation numbers of field modes within the finite region. Indeed, 
in the general definition in eq.~(\ref{mpf2}):
\begin{equation}\label{pv2dovrebbe}
\Pro_V=\sum_{h_V} \ket{h_V} \bra{h_V}
\end{equation}
the states $\ket{h_V}$ are a complete set of states of the Hilbert space of the 
localized problem $H_V$, where the degrees of freedom are values of the field in 
each point of the region $V$, i.e. $\{ \psi({\bf x}) \} \, | \, {\bf x} \in V$. 
Therefore, the above projector is a resolution of the identity of the localized 
problem and can be written as (for a real scalar field):
\begin{equation}\label{pv2}
   \Pro_V = \int_V \mathcal{D} \psi \ket{\psi}\bra{\psi}
\end{equation}
where $\ket{\psi} \equiv \otimes_{{\bf x}} \ket{\psi({\bf x})}$ and $\mathcal{D}\psi$ 
is the functional measure; the index $V$ means that the 
functional integration must be performed over the field degrees of freedom in 
the region $V$, that is ${\mathcal D} \psi = \prod_{{\bf x}\in V} d \psi({\bf x})$.
The normalization of the states is chosen so as to $\braket{\psi({\bf x})}{\psi'({\bf x})}
= \delta (\psi({\bf x})-\psi'({\bf x}))$ to ensure the idempotency of $\Pro_V$.
If we now want to give expressions like:
\begin{equation}\label{scalprod}
\bra{\Nj,\{ p \}} \Pro_V \ket{\Nj,\{ p \}}
\end{equation}
a clear meaning, we should find a way of completing the tensor product in the
projector (\ref{pv2}) with the field states outside the region $V$ such a way 
the scalar product can be performed unambiguously. 

Unfortunately the answer to this question is not unique and the projector can 
be extended to $H$ in infinitely many ways. What is important is that
the result of the calculation is independent of how the projector has been 
extended. Thus, at the end of the calculation, one has to check whether spurious 
terms appear, possibly divergent, depending explicitely on the chosen state of 
the field {\em outside} $V$ and these terms must be subtracted away. In general, 
all terms depending on the degrees of freedom of the field out of $V$ must be 
dropped from the final result.

In this work, we will extend the projector with eigenstates of the field, where
the field function $\psi({\bf x})$ is some arbitrary function outside the region 
$V$. Thus, the projector $\Pro_V$ (\ref{pv2}) is mapped to:
\begin{equation}\label{pv3}
   \Pro_V = \int_V \mathcal{D} \psi \ket{\psi}\bra{\psi} \qquad 
   \ket{\psi} \equiv \otimes_{{\bf x}\in V} \ket{\psi({\bf x})} 
   \otimes_{{\bf x}\notin V}\ket{\psi({\bf x})}
\end{equation}
where the index $V$ still implies 
${\mathcal D} \psi = \prod_{{\bf x}\in V} d \psi({\bf x})$.
We will see that, with the definition (\ref{pv3}), spurious terms depending 
on the degrees of freedom outside $V$ do arise indeed, but that they can be
subtracted ``by hand'' in a consistent way.

\section{Single particle channel}

We will start calculating the expectation value $\bra{p} \Pro_V \ket{p}$ of
a single particle channel in the simple cases of neutral and charged scalar 
fields. This is preparatory to the general case of multiparticle states in
Sect.~6.

\subsection{Neutral scalar field}

We consider a gas made of one type of spinless boson described by the free 
real scalar field\footnote{Henceforth, the capital letter $\Psi$ will denote 
field operators while for field functions we will use the small letter $\psi$.} 
(in Schr\"odinger representation):
\begin{equation}\label{scfield1}
\Psi({\bf x}) = \frac{1}{(2 \pi)^{\frac{3}{2}}} \int \frac{\d^3 \p}{\sqrt{2 \varepsilon}} 
\left( \; a(p)\; \e^{i{\bf p} \cdot {\bf x}}+a^\dagger(p)\;\e^{-i{\bf p} \cdot {\bf x}} \right)   
\end{equation} 
where $\varepsilon \equiv \sqrt{\p^2+m^2}$ is the energy, $\p$ is the 
modulus of the three-momentum and the normalization has been chosen so as to 
have the following commutation rule between annihilation and creation operators:
\begin{equation}\label{scfield2}
[a(p),a^\dagger(p')]= \delta^3({\bf p}-{\bf p}') \; .
\end{equation} 
We start writing the one-particle Fock state $\ket{p}$ in terms of creation and 
annihilation operators acting on the vacuum:
\begin{equation}\label{scfield0}
\bra{p} \Pro_V \ket{p}=\bra{0} a(p) \; \Pro_V \; a^\dagger(p)\ket{0} \; .
\end{equation} 
Since $\Pro_V$ is defined, according to (\ref{pv3}) as a functional integral of 
eigenvectors of the field operator $\Psi$, it is convenient to express creation 
and annihilation operators in terms of the field operators. We shall use following 
expressions which are the most appropriate for our task:
\begin{eqnarray}\label{scfield4}
\bra{0} a(p)&=& \bra{0} \; \frac{1}{(2 \pi)^{\frac{3}{2}}} \int \d^3 \x 
 \; \e^{-i {\bf p} \cdot {\bf x}} 
\sqrt{2 \varepsilon} \; \Psi({\bf x}) \nonumber \\ 
a^\dagger (p) \ket{0} &=& \frac{1}{(2 \pi)^{\frac{3}{2}}}\int \d^3 \x\; 
\e^{i {\bf p} \cdot {\bf x}} \sqrt{2 \varepsilon} \;\Psi ({\bf x})  \;  \ket{0} \; .
\end{eqnarray} 
These expressions can be easily checked by plugging, on the right hand side, the 
field operators in (\ref{scfield1}). By using eq.~(\ref{scfield4}) in (\ref{scfield0}):
\begin{equation}\label{scfield5}
\bra{0} a(p )\Pro_V a^{\dagger}(p )\ket{0} = 
\frac{1}{(2 \pi)^3} \int \d^3 \x \int \d^3 \x' \, \e^{i {\bf p}\cdot 
( {\bf x} -{\bf x}')} 2\varepsilon \; \bra{0} \Psi ({\bf x}')\Pro_V \Psi({\bf x})\ket{0}
\end{equation}
It can be easily checked now that the rightmost factor in the above equation turns 
out to be (by using the definition (\ref{pv3})):
\begin{equation}\label{scfield6}
\bra{0} \Psi({\bf x}')\Pro_V \Psi ({\bf x})\ket{0} =
 \int_V \mathcal{D} \psi \, 
 |\braket{\psi}{0}|^2 \psi({\bf x}') \psi({\bf x}) \; .
\end{equation}
where $\psi({\bf x})$ and $\psi({\bf x}')$ are field functions or the eigenvalues 
of the field operator relevant to the state $\ket{\psi}$, that is:
\begin{equation}
 \Psi({\bf x})\ket{\psi}=\ket{\psi}\psi({\bf x}) \; .
\end{equation}
It is possible to find a solution of the functional integral (\ref{scfield6}) by 
first considering the infinite volume limit, when the projector $\Pro_V$ reduces 
to the identity. In this limiting case, the functional integral in (\ref{pv3}) 
is now performed over all possible field functions and eq.~(\ref{pv3}) becomes 
a resolution of the identity; $\bra{0} \Psi({\bf x}')\Psi ({\bf x})\ket{0}$ is 
just the two-point correlation function that we write, according to (\ref{scfield6}):
\begin{equation} \label{scfield6a}
\bra{0} \Psi({\bf x}') \Psi ({\bf x})\ket{0} = \int \mathcal{D} \psi \;
|\braket{\psi}{0}|^2 \psi({\bf x}') \psi({\bf x}) 
\end{equation}
The product $ \braket{\psi}{0}$ is known as the {\em vacuum functional} and 
reads~\cite{weinberg}, for a scalar neutral field:
\begin{equation}\label{vacuumfunctional}
\braket{\psi}{0}  =  \mathcal{N} \exp{\left\{ - \frac{1}{4} \int \d^3 
\x_1 \int \d^3 \x_2 \, \psi({\bf x}_1) K({\bf x}_1-{\bf x}_2)\psi({\bf x}_2) 
 \right\}}
\end{equation}
where $\mathcal{N}$ is a field-independent normalization factor, which is irrelevant 
for our purposes. The function $K({\bf x}'-{\bf x})$ is called {\em kernel} and 
fulfills the equation \cite{weinberg}:
\begin{equation}\label{kernfullfills}
\int \d^3 \x' \; \e^{-i{\bf p} \cdot {\bf x}'} K({\bf x}'-{\bf x}) = 
2\varepsilon \e^{-i{\bf p} \cdot {\bf x}}
\end{equation}
whose solution is:
\begin{equation}\label{scfield11}
 K({\bf x}'-{\bf x})= \frac{1}{(2 \pi)^3} \int \d^3 \p \; \e^{-i{\bf p} 
\cdot ({\bf x}'-{\bf x})} \; 2
\varepsilon \; .
\end{equation}
The functional integral (\ref{scfield6a}) is therefore a gaussian integral and 
can be solved by using the known formulae for multiple gaussian integrals of real
variables~\cite{weinberg}:
\begin{eqnarray}\label{gauintR}
&&I_{2N}=\int \mathcal{D} \psi \; \prod_{i=1}^{2N} \psi(\xiv_i) 
\exp{\left[ - \frac{1}{2} \int \d^3 \x_1 \int \d^3 \x_2 \, \psi({\bf x}_1) 
K({\bf x}_1-{\bf x}_2)\psi({\bf x}_2) \right] }\qquad  \\ \nonumber
&& =I_0 \sum_{\stackrel{{\rm pairings }}{{\rm of} \; \xi_1,\ldots,
\xi_{2N}}} \prod_{{\rm pairs}} K^{-1}({\rm paired \; variables}) 
\end{eqnarray} 
where {\em paired variables} means couples $(\xiv_i,\xiv_j)$ whose difference
$\xiv_i-\xiv_j$ (or, what is the same, $\xiv_j-\xiv_i$ as $K^{-1}$ is symmetric)
is the argument of $K^{-1}$. The factor $I_0$ is just the normalization of the 
vacuum state $I_0=\braket{0}{0}$ which is set to 1. The inverse kernel
$K^{-1}$ can be found from its definition:
\begin{equation}\label{scfield13}
 \int\d^3 \x' \, K({\bf y}-{\bf x}')K^{-1}({\bf x}'-{\bf x})= 
 \delta^3({\bf x}-{\bf y}) \qquad \forall \; {\bf x},{\bf y}
\end{equation} 
leading to:
\begin{equation}\label{scfield110}
 K^{-1}({\bf x}'-{\bf x})= \frac{1}{(2 \pi)^3} \int \frac{\d^3 \p}{2 \varepsilon} 
 \; \e^{-i{\bf p} \cdot ({\bf x}'-{\bf x})} = 
 \bra{0} \Psi({\bf x}')\Psi ({\bf x})\ket{0}
\end{equation}
The last equality comes from (\ref{gauintR}) and (\ref{scfield6a}) in the special 
case $N=2$ or can be proved directly from field Fourier expansion (\ref{scfield1}).

We are now in a position to solve the functional integral (\ref{scfield6}) at 
finite volume. First, the functional integration variables are separated from
those which are not integrated, i.e. the field values out of $V$ ($\bar V$ 
denotes the complementary of $V$):
\begin{eqnarray}\label{scfield12}
 | \braket{\psi}{0}|^2  &=& |{\mathcal N}|^2 
  \exp{\left[ - \frac{1}{2} \int \d^3 \x_1 \int \d^3 \x_2 \, \psi({\bf x}_1) 
  K({\bf x}_1-{\bf x}_2)\psi({\bf x}_2) \right] } \nonumber \\
 &=& |{\mathcal N}|^2  \exp{\left[ - \frac{1}{2} \int_V \d^3 \x_1 \int_V \d^3 
  \x_2 \, \psi({\bf x}_1) K({\bf x}_1-{\bf x}_2)\psi({\bf x}_2) \right] } \nonumber \\
 & \times & \exp{ \left[ - \frac{1}{2} \int_{\bar V} \d^3 \x_1 \int_{\bar V} \d^3 
  \x_2 \, \psi({\bf x}_1) K({\bf x}_1-{\bf x}_2) \psi({\bf x}_2) \right] } \nonumber \\
 & \times & \exp{ \left[ - \int_V \d^3 \x_1 \int_{\bar V} \d^3 \x_2 \, 
  \psi({\bf x}_1) K({\bf x}_1-{\bf x}_2) \psi({\bf x}_2) \right] } 
\end{eqnarray}
where we have taken advantage of the symmetry of the kernel $K$. The (\ref{scfield12})
is a gaussian functional, with a general quadratic form in the field values in the
region $V$; it can be integrated in (\ref{scfield6}) according to standard rules 
\cite{weinberg}, yielding:
\begin{eqnarray}\label{funcint}
&&\int_V \mathcal{D} \psi \;| \bra{0} \psi \rangle |^2 \psi({\bf x}') \psi({\bf x})
 = K^{-1}_V({\bf x}',{\bf x}) |{\mathcal N}|^2 \det \left[ \frac{K_V}{2\pi} \right]^{-1/2} 
  \exp \left[        \right.               \nonumber \\
&& \frac{1}{2} \int_V \d^3 \x_1 \d^3 \x'_1 \int_{\bar V} \d^3 \x_2 \d^3 \x'_2 \,
  K^{-1}_V({\bf x}_1,{\bf x}'_1) K({\bf x}_1-{\bf x}_2) K ({\bf x}'_1-{\bf x}'_2)
  \psi({\bf x}'_2) \psi({\bf x}_2) \nonumber \\
&& \left. - \frac{1}{2} \int_{\bar V} \d^3 \x_1 \int_{\bar V} \d^3 \x_2 \, 
   \psi({\bf x}_1) K({\bf x}_1-{\bf x}_2) \psi({\bf x}_2) \right] \nonumber \\
&& = K^{-1}_V({\bf x}',{\bf x}) \int_V \mathcal{D} \psi \;| \bra{0} \psi \rangle |^2 =
     K^{-1}_V({\bf x}',{\bf x}) \bra{0} \Pro_V \ket{0}
\end{eqnarray}
The function $K^{-1}_V$ is the inverse of $K$ over the region $V$, namely the inverse of:
\begin{equation}\label{kappav}
K({\bf x}'-{\bf x}) \Theta_V({\bf x}')\Theta_V({\bf x}) \equiv K_V({\bf x}'-{\bf x}) \; ,
\end{equation} 
the function $\Theta_V({\bf x})$ being the Heaviside function:
\begin{eqnarray}\label{heaviside}
\Theta_V({\bf x})= \left\{ \begin{array}{ll} 1 \qquad& {\rm if}~{\bf x} \in V \\
0 & {\rm otherwise.} 
\end{array} \right.
\end{eqnarray} 
The inverse kernel $K^{-1}_V$ fulfills, by definition, the integral equation:
\begin{equation}\label{inteq1}
 \int_V \d^3 \x' \, K_V({\bf y}-{\bf x}')K_V^{-1}({\bf x}',{\bf x}) = 
 \delta^3 ({\bf x} -{\bf y})
 \qquad \forall \; {\bf x},{\bf y} \in V 
\end{equation}
Note that, because of the finite region of integration, the inverse kernel may now
depend on both the space variables instead of just their difference. Also note
that $K_V^{-1}$ is real and symmetric, being $K_V$ real and symmetric. 
Therefore, the result of the functional integration yields the simple formula:   
\begin{equation}\label{scfield14}
 \bra{0} \Psi({\bf x}')\Pro_V \Psi ({\bf x})\ket{0} = 
 K_V^{-1}({\bf x}',{\bf x}) \bra{0} \Pro_V \ket{0}
\end{equation}
where the factor $\bra{0} \Pro_V \ket{0}$ is a positive constant which we will 
leave unexpanded.

Altogether, the presence of the projector $\Pro_V$ in the eq.~(\ref{scfield14}) 
modifies the two-point correlation function by introducing a constant factor 
$\bra{0} \Pro_V \ket{0}$ and replacing the inverse kernel $K^{-1}$ with a 
different one $K_V^{-1}$. It can be easily proved, by using the general formulae
of gaussian integrals, that this holds true in the more general case of many-points
correlation function. In fact, the (\ref{gauintR}) holds for general quadratic
forms in the field $\psi$ and so the eq.~(\ref{scfield14}) can be generalized to:
\begin{equation}\label{scfield14a}
 \bra{0} \prod_{\parti=1}^{N} \Psi({\bf x}_\parti) \Pro_V \prod_{\parti=N+1}^{2N} 
 \Psi({\bf x}_\parti)\ket{0} =
 \bra{0} \Pro_V \ket{0} \!\!\!\!\!\! 
 \sum_{\stackrel{{\rm pairings }}{{\rm of} \; x_1,\ldots,
  x_{2N}}} \prod_{{\rm pairs}} K_V^{-1}({\rm paired \; var.}) \; . 
\end{equation}

The remaining task is to calculate the inverse kernel $K^{-1}_V$ by means
of (\ref{inteq1}). In fact, we will look for a solution of the more general 
equation:
\begin{equation}\label{inteq2}
 \int_V \d^3 \x' \, K ({\bf y}-{\bf x}')K_V^{-1}({\bf x}',{\bf x})= 
 \delta^3 ({\bf x} -{\bf y})
 \qquad \forall \; {\bf x} \in V, {\bf y} 
\end{equation}
with unbounded ${\bf y}$. It is clear that a solution $K^{-1}_V$ of equation
(\ref{inteq2}) is also a solution of (\ref{inteq1}) because $K_V$ equals $K$ 
when ${\bf y} \in V$. Setting ${\bf y}$ unbounded allows us to find an implicit form 
for $K^{-1}_V$. In fact (\ref{inteq2}) implies: 
\begin{equation}\label{scfield13a}
\frac{1}{(2 \pi)^3}\int_V \d^3 \x' \;  \e^{i{\bf p} \cdot {\bf x}'} \; 
 2 \varepsilon\; K^{-1}_V({\bf x}',{\bf x})=\frac{\e^{i{\bf p} \cdot {\bf x}}}
 {(2 \pi)^3}
\end{equation} 
which is obtained multiplying both sides of (\ref{inteq2}) by 
$\e^{i {\bf p}\cdot{\bf y}}/(2 \pi)^3$ and integrating over the {\em whole space} 
in $\d^3 {\rm y}$. 

We are now in a position to accomplish our task of calculating $\bra{p}\Pro_V\ket{p}$.
By plugging (\ref{scfield14}) into (\ref{scfield5}) we get:
\begin{equation}\label{scfield15}
\bra{0} a(p )\Pro_V a^{\dagger}(p )\ket{0} =\frac{1}{(2 \pi)^3} \int \d^3 \x \int 
 \d^3 \x'  \; \e^{i {\bf p}\cdot ( {\bf x} -{\bf x}')} \; 2\varepsilon \; 
 K_V^{-1}({\bf x}',{\bf x})\bra{0}\Pro_V\ket{0}  \; .
\end{equation}
The integration domain in (\ref{scfield15}) can be split into the region $V$ 
and the complementary $\bar V$ for both variables. The inverse kernel $K^{-1}_V$ 
is not defined out of $V$ and can then be set to an arbitrary value, e.g. 
zero. Otherwise, even if one chose a non vanishing prolongation of $K^{-1}_V$, an 
integration outside the domain $V$ would involve the degrees of freedom of the 
field out of $V$ and, according to the general discussion at the end of 
Sect.~4, the contributing term should be dropped. Therefore, retaining only the 
physically meaningful term, the (\ref{scfield15}) turns into:
\begin{equation}\label{scfield15a}
 \bra{0} a(p )\Pro_V a^{\dagger}(p )\ket{0} =\frac{1}{(2 \pi)^3} \int_V \d^3 \x 
 \int_V \d^3 \x' \, \e^{i {\bf p}\cdot ( {\bf x} -{\bf x}')} \; 2\varepsilon \; 
 K_V^{-1}({\bf x}',{\bf x})\bra{0}\Pro_V\ket{0}  \; .
\end{equation}
In the above equation one can easily recognize the complex conjugate of 
the left hand side of (\ref{scfield13a}). Hence, replacing it with the complex 
conjugate of the right hand side, one gets:
\begin{eqnarray}\label{scfield16}
\bra{0} a(p )\Pro_V a^{\dagger}(p )\ket{0} = 
\frac{1}{(2 \pi)^3} \int_V \d^3 \x \, \bra{0} \Pro_V \ket{0} = 
\frac{V}{(2 \pi)^3} \, \bra{0} \Pro_V \ket{0}
\end{eqnarray}
which is the same result of NRQM in (\ref{clust}) in the simple case $N=1$, 
times a factor $\bra{0} \Pro_V \ket{0}$. This factor still contains a 
dependence on the field degrees of freedom out of $V$, according to the 
projector expression in eq.~(\ref{pv3}), which should disappear at some point. 
However, we will see that this factor appears at any multiplicity and therefore 
becomes irrelevant for the calculations of the statistical averages. 

\subsection{Charged scalar field}

The calculation done for a neutral scalar field can be easily extended to a
charged scalar field. The 2-component charged scalar field in Schr\"odinger 
representation reads:
\begin{eqnarray}\label{chscfield1}
\Psi({\bf x}) = \frac{1}{(2 \pi)^{\frac{3}{2}}} \int \frac{\d^3 \p}
{\sqrt{2 \varepsilon}} \left( \; a(p)\; \e^{i{\bf p} \cdot {\bf x}}+
b^\dagger(p)\;\e^{-i {\bf p} \cdot {\bf x}} \right) \\ \nonumber
\Psi^\dagger({\bf x}) = \frac{1}{(2 \pi)^{\frac{3}{2}}} 
\int \frac{\d^3 \p}{\sqrt{2 \varepsilon}} \left( \; b(p)\; 
\e^{i{\bf p} \cdot {\bf x}}+a^\dagger(p)\;\e^{-i{\bf p} \cdot {\bf x}} \right) 
\end{eqnarray} 
where $a$, $a^\dagger$ and $b$, $b^\dagger$ are annihilation and creation operators 
of particles and antiparticles respectively. They satisfy commutation relations:
\begin{eqnarray}\label{chscfield2}
[a(p),a^\dagger(p')] &=& [b(p),b^\dagger(p')]=
\delta^3({\bf p}-{\bf p}') \\ \nonumber
[a(p),b(p')]&=& [a^\dagger(p),b(p')]=0
\end{eqnarray} 
Likewise, the fields obey the commutation relations:
\begin{equation}\label{comchfield}
[\Psi({\bf x}),\Psi^\dagger({\bf y})]=0 \; .
\end{equation} 
and it is then possible to construct field states $\ket{\psi,\psi^\dagger}$. The 
projector $\Pro_V$ can be written as: 
\begin{equation}\label{pv4}
\Pro_V \equiv \int_V \mathcal{D} (\psi^\dagger,\psi) 
\ket{\psi,\psi^\dagger}\bra{\psi, \psi^\dagger} 
\end{equation}
with suitable state normalization and arbitrary field functions $\psi(\bf{x})$
out of the region $V$ \footnote{The functional measure in the equation (\ref{pv4})
reads $\prod_{\bf x} d\psi({\bf x})d\psi^*({\bf x})/i\pi$. Anyhow, its explicit
form is not important for our purposes.} 
Similarly to eq.~(\ref{scfield4}), one can write:
\begin{eqnarray}\label{chscfield3}
\bra{0} a(p)&=& \bra{0} \;\frac{1}{(2 \pi)^{\frac{3}{2}}} \int \d^3 \x 
\; \e^{-i {\bf p} \cdot {\bf x}} \sqrt{2 \varepsilon} \; \Psi({\bf x})  
\\ \nonumber
a^\dagger (p) \ket{0} &=& \frac{1}{(2 \pi)^{\frac{3}{2}}} \int \d^3 \x
\; \e^{i {\bf p} \cdot {\bf x}} \sqrt{2 \varepsilon} \; \Psi^\dagger ({\bf x}) 
\; \ket{0} \\ \nonumber
\bra{0} b(p )&=& \bra{0} \;\frac{1}{(2 \pi)^{\frac{3}{2}}} \int \d^3 \x 
\; \e^{-i {\bf p} \cdot {\bf x}} \sqrt{2 \varepsilon} \; \Psi^\dagger ({\bf x})   
\\  \nonumber
b^\dagger (p ) \ket{0} &=& \frac{1}{(2 \pi)^{\frac{3}{2}}} \int \d^3 \x 
\; \e^{i {\bf p} \cdot {\bf x}} \sqrt{2 \varepsilon} \;
 \Psi({\bf x})\;\ket{0} \; 
\end{eqnarray}
The chain of arguments of the previous subsection can be repeated and the 
functional integral:
\begin{equation}\label{chscfield3a}
\bra{0} \prod_{\parti=1}^{N} \Psi({\bf x}_\parti) \Pro_V \prod_{\parti=1}^{N} 
\Psi^\dagger({\bf x}'_\parti)\ket{0} =
\int_V \mathcal{D} (\psi^\dagger,\psi)  \;| \bra{0} \psi ,\psi^\dagger  
\rangle |^2 \prod_{\parti=1}^{N} \psi({\bf x}_\parti)\psi^\dagger({\bf x}'_\parti)  
\end{equation}
found to be a multiple gaussian integral. Letting $\rho$ be a permutation of the 
integers $1,\ldots,N$, the integration on the right hand side of eq.~(\ref{chscfield3a}) 
yields:
\begin{equation}\label{chscfield14a}
\bra{0} \prod_{\parti=1}^{N} \Psi({\bf x}_\parti) \Pro_V \prod_{\parti=1}^{N} 
\Psi^\dagger({\bf x}'_\parti)\ket{0} =
\bra{0} \Pro_V \ket{0}\sum_{\rho \in {\rm S}_N} 
\prod_{\parti=1}^N K_V^{-1}({\bf x}_\parti,{\bf x}'_{\rho(\parti)})   
\end{equation}
which differs from the corresponding expression for the real scalar field because 
now the field is complex and $\psi$ can only be coupled to $\psi^\dagger$ 
\cite{stone}:
\begin{eqnarray}\label{gauintC}
\!\!\!\!\!\!\!\!\!\!\!\!&&\int \mathcal{D} (\psi^\dagger,\psi) \, 
\prod_{\parti=1}^{N} \left( \psi(\xiv_\parti) \psi^\dagger(\xiv_\parti') \right)
\exp\left[ -\int \d^3 \x_1 \int \d^3 \x_2 \, \psi^\dagger({\bf x}_1) 
K({\bf x}_1-{\bf x}_2) \psi({\bf x}_2)\right] \nonumber \\ 
\!\!\!\!\!\!\!\!\!\!\!\!&& = I_0 \sum_{\rho \in {\rm S}_N} \prod_{\parti=1}^{N} 
K^{-1}(\xiv_\parti-\xiv_{\rho(\parti)}') 
\end{eqnarray} 
However, the functional integral involving only two fields $\psi$ and $\psi^\dagger$
yields the same result as for neutral particles. Thus, the kernel $K$ is still the 
same and so is the integral 
equation (\ref{scfield13a}) defining $K_V^{-1}$. The expectation value of $\Pro_V$
in a state with only one particle (or antiparticle) will also be the same as in 
eq.~(\ref{scfield16}), that is:
\begin{equation}\label{chscfinal}
\bra{0} a(p )\Pro_V a^{\dagger}(p )\ket{0} = 
\bra{0} b(p )\Pro_V b^{\dagger}(p )\ket{0}=\frac{V}{(2 \pi)^3}\bra{0} \Pro_V \ket{0}
 \; .
\end{equation}
%

\section{Multiparticle channels}

We have seen in the previous section that the expectation value 
$\bra{p}\Pro_V \ket{p}$ for a single spinless particle is the same obtained 
in a NRQM approach \cite{bf1} times an overall immaterial factor 
$\bra{0} \Pro_V \ket{0}$. In this section, we will tackle the calculation of 
the general multiparticle state. We will see that, by using the projector definitions
in eqs.~(\ref{pv3}),(\ref{pv4}) and subtracting the spurious contributions 
stemming from external field degrees of freedom, the final result is still 
the same as in the NRQM calculation times the factor $\bra{0} \Pro_V \ket{0}$.

We will first address the case of $N$ charged particles.

\subsection{Identical charged particles}

We will consider a state with $N$ identical charged particles; for $N$ antiparticles
the result is trivially the same. 

We want to calculate:
\begin{equation}\label{nparticles}
\bra{N,\{p \}} \Pro_V \ket{N,\{ p \}} = 
\bra{0} \prod_{\parti}^N a(p_\parti) \Pro_V \prod_{\parti}^N a^\dagger(p_\parti)\ket{0} \; .
\end{equation}
Since:
\begin{equation}\label{zerocomm}
[a^\dagger, \Psi^\dagger]=[a,\Psi]=[b,\Psi^\dagger]=[b^\dagger,\Psi]=0
\end{equation}
one can replace creation and annihilation operators with their expressions in 
(\ref{chscfield3}) disregarding the position of the operators with respect to the 
vacuum state. In formula:
\begin{eqnarray}\label{nparticles2}
\bra{N,\{ p \}} \Pro_V \ket{N,\{ p \}} &=& 
\prod_{\parti=1}^N \left[ \frac{1}{(2 \pi)^3} \int \d^3 \x_\parti \int
 \d^3 \x'_\parti \; \e^{-i {\bf p}_\parti \cdot ({\bf x}_\parti-{\bf x}'_\parti)} 
 2 \varepsilon_\parti \right] \\ \nonumber 
&\times& \bra{0} \prod_{\parti=1}^N \Psi({\bf x}_\parti) \Pro_V 
 \prod_{\parti=1}^N \Psi^\dagger ({\bf x}'_\parti) \ket{0} \\  \nonumber 
&=&  \prod_{\parti=1}^N \left[\frac{1}{(2 \pi)^3} \int \d^3 \x_\parti \int
\d^3 \x'_\parti \; \e^{-i {\bf p}_\parti \cdot ({\bf x}_\parti-{\bf x}'_\parti)} 
2 \varepsilon_\parti \right] \nonumber \\ 
&\times& \sum_{\rho \in {\rm S}_N} \prod_{\parti=1}^N 
K_V^{-1}({\bf x}_\parti,{\bf x}'_{\rho(\parti)}) \bra{0} \Pro_V \ket{0} \; ,
\end{eqnarray}
where ${\rm S}_N$ is the permutation group of rank $N$. Now, like for the single
particle case, we restrict the integration domain to $V$ in (\ref{nparticles2}) 
in order to get rid of external degrees of freedom and, by repeatedly using 
eq.~(\ref{scfield13a}), we are left with:
\begin{eqnarray}\label{nparticles3}
\frac{\bra{N,\{ p \}} \Pro_V \ket{N,\{ p \}}}{\bra{0}  
\Pro_V \ket{0}} &=& 
\sum_{\rho \in {\rm S}_N} \prod_{\parti=1}^N \left[\frac{ 1}{(2 \pi)^3} \int_V  
\d^3 \x'_\parti \; 
\e^{-i {\bf p}_\parti \cdot ({\bf x}'_{\rho(\parti)}-{\bf x}'_\parti)}  \right] \nonumber \\
&=& \sum_{\rho \in {\rm S}_N} \prod_{\parti=1}^N \left[\frac{1}{(2 \pi)^3} \int_V  
\d^3 \x'_\parti \; \e^{-i ({\bf p}_{\rho^{-1}(n)}-{\bf p}_n)\cdot {\bf x}'_n} \right] \nonumber \\
\end{eqnarray}
Hence, using the definition (\ref{fourier2}):
\begin{equation}\label{nparticles4}
\bra{N, \{ p \}} \Pro_V \ket{N,\{ p \}} = \sum_{\rho \in {\rm S}_N} 
 \prod_{\parti=1}^N  F_V({\bf p}_{\rho(\parti)}-{\bf p}_\parti) \bra{0}\Pro_V\ket{0}
\end{equation}
which is exactly the expression~(\ref{clust}) obtained in NRQM for $N$ identical 
bosons, times the factor $\bra{0} \Pro_V \ket{0}$.

\subsection{Identical neutral particles}

The case of $N$ identical neutral particles is more complicated because of the
possibility of particle pairs creation. This is reflected in the formalism in
the occurrence of many additional terms in working out the eq.~(\ref{nparticles}).
We start with the calculation for a state with two neutral particles of four-momenta 
$p_1$ and $p_2$ respectively, generalizing to $N$ particles thereafter. In terms of 
creation and annihilation operators:
\begin{equation}\label{scfield0a}
\bra{p_1,p_2} \Pro_V \ket{p_1,p_2}=\bra{0} a(p_1) a(p_2)\; \Pro_V \;a^\dagger(p_2) 
a^\dagger(p_1)\ket{0} 
\end{equation} 
which we can rewrite using eq.~(\ref{scfield4}) for $a(p_1)$ and $a^\dagger(p_1)$ as:
\begin{eqnarray}\label{scfield18}
\bra{p_1,p_2} \Pro_V \ket{p_1,p_2}&=& \frac{1}{(2 \pi)^3} \int \d^3 \x_1 \int 
\d^3 \x_1' \, \e^{-i {\bf p}_1\cdot ( {\bf x_1} -{\bf x_1}')}  \nonumber \\
&\times& 2\varepsilon_1\bra{0} \Psi ({\bf x}_1) a(p_2)\; \Pro_V \;a^\dagger(p_2) 
\Psi ({\bf x}_1')\ket{0} \; .
\end{eqnarray} 
In order to use the expression (\ref{scfield4}) for annihilation and creation
$a(p_2)$ and $a^\dagger(p_2)$ operators we have to get them acting on the vacuum state, 
hence they should be moved from their position in eq.~(\ref{scfield18}) outwards. 
This can be done by taking advantage of the following commutation rules:
\begin{eqnarray}\label{scfield19}
&&[\Psi ({\bf x}),a(p)] = - \frac{\e^{-i{\bf p} \cdot {\bf x}}}{(2 \pi)^{\frac{3}{2}}
\sqrt{2 \varepsilon} }  \nonumber \\
&& [a(p)^\dagger, \Psi ({\bf x})] = - \frac{\e^{i{\bf p} \cdot {\bf x}}}
{(2 \pi)^{\frac{3}{2}}\sqrt{2 \varepsilon}}
\end{eqnarray} 
Using (\ref{scfield19}) and then plugging (\ref{scfield4}) in eq.~(\ref{scfield18})
we get:
\begin{eqnarray}\label{scfield21}
\!\!\!\!\!\!\!\!&&\bra{p_1,p_2} \Pro_V \ket{p_1,p_2}=  \nonumber \\
\!\!\!\!\!\!\!\!&&= \frac{\varepsilon_1}{\varepsilon_2(2 \pi)^6} \int \d^3 \x_1 \int \d^3 \x_1' \; 
 \e^{-i {\bf p}_1\cdot ( {\bf x_1} -{\bf x_1}')} 
 \e^{-i{\bf p}_2 \cdot ({\bf x}_1-{\bf x}_1')} 
\bra{0} \Pro_V \ket{0}    \nonumber \\
\!\!\!\!\!\!\!\!&& - \frac{2\varepsilon_1}{(2 \pi)^6} \int \d^3 \x_1 \int \d^3 \x_1'\int \d^3 \x_2 \, 
\e^{-i {\bf p}_1\cdot ( {\bf x_1} -{\bf x_1}')}
\e^{-i{\bf p}_2 \cdot ({\bf x}_2-{\bf x}_1')} \,  
\bra{0} \Psi ({\bf x}_2)  \Pro_V \Psi ({\bf x}_1) \ket{0}  \nonumber \\
\!\!\!\!\!\!\!\!&& - \frac{2\varepsilon_1}{(2 \pi)^6} \int \d^3 \x_1 \int \d^3 \x_1'\int \d^3 \x_2' \, 
\e^{-i {\bf p}_1\cdot ( {\bf x_1} -{\bf x_1}')}
\e^{-i{\bf p}_2 \cdot ({\bf x}_1-{\bf x}_2')}\,  
\bra{0} \Psi ({\bf x}_1')\Pro_V   \Psi ({\bf x}_2')\ket{0}  \nonumber \\
\!\!\!\!\!\!\!\!&& +\frac{4 \varepsilon_1 \varepsilon_2}{(2 \pi)^6} \int \d^3 \x_1 \int \d^3  \x_1' 
 \int \d^3 \x_2 \int \d^3 \x_2' \, \e^{-i {\bf p}_1\cdot ( {\bf x_1} -{\bf x_1}')}
\, \e^{-i {\bf p}_2\cdot ( {\bf x_2} -{\bf x_2}')} \, \nonumber \\  
\!\!\!\!\!\!\!\!&&\times \bra{0} \Psi ({\bf x}_2) \Psi ({\bf x}_1)\, \Pro_V \,
\Psi ({\bf x}_1') \Psi ({\bf x}_2') \ket{0}
\end{eqnarray} 
where advantage has been taken of the fact that $[\Pro_V,\Psi({\bf x})]=0$.
The eq.~(\ref{scfield21}) has four different terms, among which only the last 
was present in the case of 2 charged particles in Subsect.~6.1. Instead, the 
first three terms arise from contractions of the annihilation and creation operators 
with field operators on the same side with respect to $\Pro_V$. By using 
(\ref{scfield14a}) and replacing all unbounded integrations with integrations over 
$V$ to eliminate spurious degrees of freedom of the field, the eq.~(\ref{scfield21}) 
yields:
\begin{eqnarray}\label{scfield22}
&&\frac{\bra{p_1,p_2} \Pro_V \ket{p_1,p_2}}{\bra{0} \Pro_V \ket{0}}=  
\frac{\varepsilon_1}{\varepsilon_2(2 \pi)^6} \int_V \d^3 \x_1 \int_V \d^3 \x_1' \, 
\e^{-i {\bf p}_1\cdot ( {\bf x_1} -{\bf x_1}')} 
 \e^{-i{\bf p}_2 \cdot ({\bf x}_1-{\bf x}_1')} \nonumber \\
&& - \frac{2\varepsilon_1}{(2 \pi)^6} \int_V \d^3 \x_1 \int_V \d^3 \x_2  
\int_V \d^3 \x_1' \, \e^{-i {\bf p}_1\cdot ( {\bf x_1} -{\bf x_1}')}
\e^{-i{\bf p}_2 \cdot ({\bf x}_2-{\bf x}_1')}\, 
K_V^{-1}({\bf x}_2,{\bf x}_1)  \nonumber \\
&& -\frac{2\varepsilon_1}{(2 \pi)^6}\int_V \d^3 \x_1\int_V \d^3 \x_1'\int_V \d^3 \x_2' \, 
\e^{-i {\bf p}_1\cdot ( {\bf x_1} -{\bf x_1}')}
\e^{-i{\bf p}_2 \cdot ({\bf x}_1-{\bf x}_2')}\,  
K_V^{-1}({\bf x}_1',{\bf x}_2')  \nonumber  \\
&& +\frac{4 \varepsilon_1 \varepsilon_2 }{(2 \pi)^6} \int_V \d^3 \x_1 \int_V \d^3 \x_1' 
\int_V \d^3 \x_2 \int_V \d^3 \x_2' \, 
\e^{-i {\bf p}_1 \cdot( {\bf x_1} -{\bf x_1}')}\, 
\e^{-i {\bf p}_2\cdot ( {\bf x_2} -{\bf x_2}')}\,  \nonumber \\
&&\times \left[ K_V^{-1}({\bf x}_1,{\bf x}_1')K_V^{-1}({\bf x}_2,{\bf x}_2')+
K_V^{-1}({\bf x}_1,{\bf x}_2')K_V^{-1}({\bf x}_2,{\bf x}_1')  \right. \nonumber \\
&& \left.+ K_V^{-1}({\bf x}_1,{\bf x}_2)K_V^{-1}({\bf x}_2',{\bf x}_1') \right] \; .
\end{eqnarray} 
We can now use (\ref{scfield13a}) to integrate away the inverse kernel in 
eq.~(\ref{scfield22}). For the second term on the right hand side, we choose to 
integrate $\exp(-\i {\bf p}_2\cdot{\bf x}_2) K_V^{-1}({\bf x}_2,{\bf x}_1)$ in $\x_2$
and for the third term $\exp(\i {\bf p}_1\cdot{\bf x}'_1) K_V^{-1}({\bf x}'_1,{\bf x}'_2)$
in $\x'_1$. At the same time, we perform the integration of the last term in an
arbitrary couple of variables with indices 1 and 2, obtaining:
\begin{eqnarray}\label{scfield22a}
&&\frac{\bra{p_1,p_2} \Pro_V \ket{p_1,p_2}}{\bra{0} \Pro_V \ket{0}}=  
\frac{1}{(2 \pi)^6} \int_V \d^3 \x_1 \int_V \d^3 \x_1' \, 
\e^{-i {\bf p}_1\cdot ( {\bf x_1} -{\bf x_1}')} \e^{-i{\bf p}_2 \cdot ({\bf x}_1-{\bf x}_1')} 
\frac{\varepsilon_1}{\varepsilon_2} \nonumber \\
&& - \frac{1}{(2 \pi)^6} \int_V \d^3 \x_1 \int_V \d^3 \x_1' \, 
\e^{-i {\bf p}_1\cdot ( {\bf x_1} -{\bf x_1}')} 2\varepsilon_1 
\frac{\e^{-i{\bf p}_2 \cdot ({\bf x}_1-{\bf x}_1')}}{2 \varepsilon_2} \nonumber \\
&& - \frac{1}{(2 \pi)^6} \int_V \d^3 \x_1 \int_V \d^3 \x_2' \, 
\e^{-i {\bf p}_1\cdot ( {\bf x_1} -{\bf x_2}')} \e^{-i{\bf p}_2 \cdot 
({\bf x}_1-{\bf x}_2')} \nonumber \\ 
&& + \frac{V^2}{(2 \pi)^6} + \frac{1}{(2 \pi)^6}\int_V \d^3 x_1 \int_V \d^3 \x_2 \,
 \e^{-i ({\bf p_1} -{\bf p_2}) \cdot ({\bf x_1} -{\bf x_2})} \nonumber \\
&& + \frac{1}{(2 \pi)^6} \int_V \d^3 \x_1 \int_V \d^3 \x_2' \, 
\e^{-i {\bf p}_1\cdot ( {\bf x_1} -{\bf x_2}')} \e^{-i{\bf p}_2 \cdot 
({\bf x}_1-{\bf x}_2')} 
\end{eqnarray} 
Four terms in the above sum cancel out and we are left with:
\begin{equation}\label{scfield23}
\frac{\bra{p_1,p_2} \Pro_V \ket{p_1,p_2}}{\bra{0} \Pro_V \ket{0}} = 
\frac{V^2}{(2 \pi)^6}+\mid F_V({\bf p}_1-{\bf p}_2) \mid^2 
=\sum_{\rho \in {\rm S}_2} \prod_{\parti=1}^2 F_V({\bf p}_{\rho(\parti)}-{\bf p}_\parti) 
\end{equation} 
This is the same result one would obtain for two identical bosons in a NRQM 
framework \cite{bf1}. The second term on the right hand side accounts for the 
well known phenomenon of Bose-Einstein correlation in the emission of identical 
boson pairs. 

Looking back at the whole derivation, we find that the terms arising from pairings 
of field variables on the same side with respect to $\Pro_V$ have cancelled with 
the terms stemming from commutation of annihilation and creation operators with field 
operators; the only surviving terms are $N!$ pairings of field variables
on different sides of $\Pro_V$, just like in the case of charged particles. This 
cancellation property holds and so, the formula (\ref{nparticles4}) applies to the case of 
$N$ neutral particles as well. A proof based on the form of the thermodynamic 
limit $V \to \infty$ is given in Appendix B.

\subsection{Particle-antiparticle case}
\label{papapairs}

For a state with one particle and one antiparticle, the expectation value of $\Pro_V$ 
reads:
\begin{equation}
\bra{0} a(p_1) b(p_2)\; \Pro_V \;b^\dagger(p_2) a^\dagger(p_1)\ket{0} 
\end{equation} 
implying, in view of (\ref{chscfield3}):
\begin{eqnarray}\label{chscfield4}
\!\!\!\!\!\!\!\!\!
 &&\bra{0} a(p_1) b(p_2)\; \Pro_V \;b^\dagger(p_2) a^\dagger(p_1)\ket{0} 
 \nonumber \\
\!\!\!\!\!\!\!\!\!
&&=\frac{1}{(2 \pi)^3} \int \d^3 \x_1 \int \d^3 \x_1' \, 
\e^{-i {\bf p}_1\cdot ( {\bf x_1} -{\bf x_1}')}
2\varepsilon_1 \bra{0} \Psi ({\bf x}_1) b(p_2)\, \Pro_V \;b^\dagger(p_2) 
\Psi^\dagger ({\bf x}_1')\ket{0} \; .
\end{eqnarray} 
Like for neutral particles, the $b$ and $b^\dagger$ operators are moved outwards 
to get them acting on the vacuum state by using the commutators:
\begin{eqnarray}\label{chscfield5}
 && [\Psi ({\bf x}),b(p)] = \frac{\e^{-i{\bf p} \cdot {\bf x}}}
 {(2 \pi)^{\frac{3}{2}}\sqrt{2 \varepsilon} } \nonumber \\ 
 && [b(p)^\dagger, \Psi^\dagger ({\bf x})] = 
 \frac{\e^{i{\bf p} \cdot {\bf x}}}{(2 \pi)^{\frac{3}{2}}\sqrt{2 \varepsilon}}
\end{eqnarray} 
By using the eq.~(\ref{chscfield3}) and, like in the previous subsection, restricting
the integration to the region $V$ to eliminate external degrees of freedom, we obtain:
\begin{eqnarray}\label{chscfield6}
\!\!\!\!\!\!\!\!\!\!\!
&&\bra{0} a(p_1) b(p_2)\; \Pro_V \;b^\dagger(p_2) a^\dagger(p_1)\ket{0} \nonumber \\ 
\!\!\!\!\!\!\!\!\!\!\!
&& = \frac{\varepsilon_1}{\varepsilon_2 (2 \pi)^6} \int_V \d^3 \x_1 \int_V \d^3 \x_1' \, 
 \e^{-i {\bf p}_1\cdot ( {\bf x_1} -{\bf x_1}')} 
 \e^{-i{\bf p}_2 \cdot ({\bf x}_1-{\bf x}_1')} \bra{0} \Pro_V \ket{0}  \nonumber \\
\!\!\!\!\!\!\!\!\!\!\!
&& - \frac{2\varepsilon_1}{(2 \pi)^6} \int_V \d^3 \x_1 \int_V \d^3 \x_1' \int_V \d^3 \x_2 \, 
\e^{-i {\bf p}_1\cdot ( {\bf x_1} -{\bf x_1}')}
\e^{-i{\bf p}_2 \cdot ({\bf x}_2-{\bf x}_1')} \,
\bra{0} \Psi^\dagger ({\bf x}_2)  \Pro_V \Psi ({\bf x}_1)\ket{0}  \nonumber \\
\!\!\!\!\!\!\!\!\!\!\!
&& -\frac{2\varepsilon_1}{(2 \pi)^6} \int_V \d^3 \x_1 \int_V \d^3 \x_1' \int_V \d^3 \x_2'  
 \e^{-i {\bf p}_1\cdot ( {\bf x_1} -{\bf x_1}')}
 \e^{-i{\bf p}_2 \cdot ({\bf x}_1-{\bf x}_2')}\, \bra{0}  
\Psi^\dagger ({\bf x}_1') \Pro_V \Psi ({\bf x}_2')\ket{0}  \nonumber \\
\!\!\!\!\!\!\!\!\!\!\!
&& +\frac{4 \varepsilon_1 \varepsilon_2}{(2 \pi)^6} 
 \int_V \d^3 \x_1 \int_V \d^3 \x_1'\int_V \d^3 \x_2 
 \int_V \d^3 \x_2' \; \e^{-i {\bf p}_1\cdot ( {\bf x_1} -{\bf x_1}')}
 \, \e^{-i {\bf p}_2\cdot ( {\bf x_2} -{\bf x_2}')} \, \nonumber \\
\!\!\!\!\!\!\!\!\!\!\!
&&\times \bra{0} \Psi^\dagger ({\bf x}_2) \Psi ({\bf x}_1)\, \Pro_V \, 
\Psi^\dagger ({\bf x}_1') \Psi ({\bf x}_2') \ket{0} 
\end{eqnarray} 
whose result is (\ref{scfield22}) except the (last-1)th term, because now: 
\begin{eqnarray}\label{chscfield7}
&&\frac{ \bra{0} \Psi^\dagger ({\bf x}_2) \Psi ({\bf x}_1)\; \Pro_V 
\;\Psi^\dagger ({\bf x}_1') \Psi ({\bf x}_2') \ket{0}}{\bra{0}  \Pro_V \ket{0}} 
\\ \nonumber
&&=K_V^{-1}({\bf x}_1,{\bf x}_1')K_V^{-1}({\bf x}_2,{\bf x}_2')+ 
K_V^{-1}({\bf x}_1,{\bf x}_2)K_V^{-1}({\bf x}_2',{\bf x}_1') 
\end{eqnarray} 
which is a consequence of the general expression (\ref{chscfield14a}).
Again, four out of five terms cancel out in eq.~(\ref{chscfield6}) and we get:
\begin{equation}\label{chscfield24}
\frac{\bra{0} a(p_1) b(p_2)\; \Pro_V \;b^\dagger(p_2) 
a^\dagger(p_1)\ket{0}}{\bra{0} \Pro_V \ket{0}}= \frac{V^2}{(2 \pi)^6}
\end{equation} 
where the missing term with respect to the neutral particle case (\ref{scfield23}) 
is the one involving permutations; this is a natural result because particles and 
antiparticles are, of course, not identical. Hence, the expectation value of $\Pro_V$ 
on a particle-antiparticle pair shows a remarkable factorization property, that is:
\begin{equation}\label{chscfield0a}
  \bra{0} a(p_1) b(p_2)\, \Pro_V \, b^\dagger(p_2) a^\dagger(p_1)\ket{0} =
  \bra{0} a(p_1) \Pro_V a^\dagger(p_1)\ket{0} \bra{0} b(p_2) \Pro_V b^\dagger(p_2)\ket{0}
\end{equation} 
where the $=$ sign holds provided that the integrations are restricted to the
region $V$.

Extrapolating to the most general case of $N_+$ particles and $N_-$ antiparticles,
it can be argued by using the limit $V \to \infty$ that the factorization of 
the microcanoncal state weight holds (see Appendix B) at any multiplicity, i.e. 
particles and antiparticles behave like two different species:
\begin{eqnarray}\label{twospecies}
 && \bra{\{N_+\},\{p_+\},\{N_-\},\{p_-\}} \Pro_V \ket{\{N_+\},\{p_+\},\{N_-\},\{p_-\}}
  \nonumber \\
 &=& \sum_{\rho_+ \in {\rm S}_{N_+}} \prod_{n_+ = 1}^{N_+} 
   F_V ({\bf p}_{\rho_+(n_+)}-{\bf p}_{n_+}) 
  \sum_{\rho_- \in {\rm S}_{N_-}} \prod_{n_- = 1}^{N_-} 
   F_V ({\bf p}_{\rho_-(n_-)}-{\bf p}_{n_-}) \bra{0}\Pro_V \ket{0} \; .
\end{eqnarray}
%

\section{Summary and discussion}
\label{sec:discussion} 

On the basis of eqs.~(\ref{multexp}), (\ref{multexp2}), (\ref{t4}) and 
(\ref{nparticles4}), which applies to charged as well as neutral particles, and 
taking into account that the states $\ket{N,\{p\}}$ have been chosen to be 
eigenvectors of four-momentum, we can finally write down the full expression 
of the microcanonical partition function of a relativistic quantum gas of neutral 
spinless bosons as:
\begin{eqnarray}\label{mpffinal}
 \Omega &=& \bra{0} \Pro_V \ket{0} 
 \sum_{N=0}^\infty \frac{1}{N!}\left[ \prod_{n=1}^{N} \int \d^3 \p_{n} 
 \right]\delta^4 \left( P - \sum_{n=1}^{N} p_n \right) \nonumber \\
 &\times& \sum_{\rho \in {\rm S}_{N}} \prod_{n=1}^{N} \int_V \d^3 \x 
 \exp[i{\bf x} \cdot ({\bf p}_{\rho(n)}-{\bf p}_{n})]
\end{eqnarray} 
with $P=(M,{\bf 0})$ and the factor $1/N!$ has been introduced in order to avoid 
multiple counting 
when integrating over particle momenta. Similarly, on the basis of eqs.~(\ref{sumovchann})
(\ref{mpf2summ}) and (\ref{twospecies}), the microcanonical partition function of 
a relativistic quantum gas of charged spinless bosons can be written as:
\begin{eqnarray}\label{mpffinal2}
\Omega &=& \bra{0} \Pro_V \ket{0} 
 \sum_{N_+,N_-=0}^\infty \frac{1}{N_+!N_-!} \left[ \prod_{n=1}^{N_++N_-} 
 \int \d^3 \p_{n} \right] \delta^4 \left( P - \sum_{n=1}^{N_+ + N_-} p_n \right) 
  \nonumber \\
 &\times& \sum_{\rho_+ \in {\rm S}_{N_+}} \prod_{n_+ = 1}^{N_+} 
   \int_V \d^3 \x \exp[i{\bf x} \cdot ({\bf p}_{\rho_+(n_+)}-{\bf p}_{n_+})]
  \nonumber \\
 &\times& \sum_{\rho_- \in {\rm S}_{N_-}} \prod_{n_- = 1}^{N_-} 
  \int_V \d^3 \x \exp[i{\bf x} \cdot ({\bf p}_{\rho_-(n_-)}-{\bf p}_{n_-})] \; .
\end{eqnarray} 
The generalization to a multi-species gas of spinless bosons is then easily 
achieved:
\begin{eqnarray}\label{mpfgeneral}
\Omega &=& \bra{0} \Pro_V \ket{0} 
 \sum_{\Nj} \left[\prod_{j=1}^{K}\frac{1}{N_j!} \prod_{\parti_j=1}^{N_j} 
 \int \d^3 \p_{\parti_j}  \right] 
 \delta^4 \left( P - \sum_{\parti=1}^{N} p_{\parti} \right) \nonumber \\
 &\times& \prod_{j=1}^{k}\sum_{\rho_j \in \; {\rm S}_{N_j}}\prod_{\parti_j=1}^{N_j} 
 \int_V \d^3 \x \exp[i{\bf x} \cdot ({\bf p}_{\rho_j({\parti_j})}-{\bf p}_{\parti_j})] \; .
\end{eqnarray} 
where $N=\sum_j N_j$. The formulae (\ref{mpffinal}), (\ref{mpffinal2}) and
(\ref{mpfgeneral}) are our final result. The finite volume Fourier 
integrals in the above expressions nicely account for quantum statistics correlations 
known as Bose-Einstein and Fermi-Dirac correlations. We stress once more that for 
a charged quantum gas, particles and antiparticles can be handled as belonging 
to distinct species and they correspond to different labels $j$ in the multi-species 
generalization of (\ref{mpfgeneral}).

The expression of the MPF (\ref{mpffinal}) is the same as obtained in a NRQM 
calculation in ref.~\cite{bf1}, quoted in this work in (\ref{multispecies}), times
an overall factor $\bra{0} \Pro_V \ket{0}$ which is immaterial for the calculation
of statistical averages in the microcanonical ensemble. More specifically, the
expectation value of $\Pro_V$ on a free multi-particle state (see eqs.~
(\ref{nparticles4}) and (\ref{twospecies})) is the same as in the NRQM calculation
(\ref{clust}) times $\bra{0} \Pro_V \ket{0}$. This result has been 
achieved enforcing a subtraction prescription, namely that all terms depending on the
degrees of freedom outside the system region $V$ must be subtracted ``by hand" in
all terms at fixed multiplicities. The factor $\bra{0} \Pro_V \ket{0}$ is still 
dependent on those spurious degrees of freedom, according to the $\Pro_V$ definition
in eq.~(\ref{pv3}), but this does not affect any statistical average because it
always cancels out. In the thermodynamic limit, this factor tends to 1 as 
$\Pro_V \to {\sf I}$ and the large volume limit result known in literature \cite{chh}
(i.e. eq.~(\ref{omold})) is recovered.

This result looks surprising in a sense because one would have expected,
{\em a priori} that quantum relativistic effects would affect the statistical averages 
bringing in a dependence on the ratio between the Compton wavelength and the linear 
size of the region, getting negligible at small values, i.e. when $V \to \infty$. 
This is because the condition (\ref{condition}) from which the MPF expressions 
(\ref{singlespecies},\ref{multispecies}) ensue, no longer holds in quantum 
field theory at finite volume and only applies with good approximation at very 
large volumes, as discussed in Sect.~4. However, in the calculation of statistical 
averages, only the projector $\Pro_V$ enters and this implies the summation over 
all states at different integrated occupation numbers $\ket{\widetilde{N}}_V$, 
according to (\ref{pv1}). So, even though the coefficients in the expansion 
(\ref{confstates}) are different from zero and do depend on the volume, it turns 
out that summing all of them at fixed $N$, one gets the same result as in 
the NRQM approximation (\ref{condition}). In formula, by using (\ref{pv1}):
\begin{eqnarray}
   && \bra{N,\{p\}}\Pro_V\ket{N\{p\}} = \sum_{\widetilde{N},\{k_V\}} 
   |\braket{N,\{p\}}{\widetilde{N},\{k_V\}}|^2 \nonumber \\
    &=& \sum_{\widetilde{N},\{k_V\}} |\braket{0}{\widetilde{N},\{k_V\}}|^2
   \sum_{\{k_V\}} |\braket{N,\{p\}}{N,\{k_V\}}|^2_{NR}
\end{eqnarray}
where the index NR means that one should make a non-relativistic quantum mechanical 
calculation.

This calculation can be extended to the most general case of the 
microcanonical ensemble of an ideal relativistic quantum gas by fixing the maximal
set of observables of the Poincar\'e algebra, i.e. spin, third component and parity
besides the energy-momentum four-vector. This will be the subject of a forthcoming
publication.

\section*{Acknowledgments}

We are grateful to F. Colomo and L. Lusanna for stimulating discussions.

 \appendix
\section{Bogoliubov relations for a real scalar field}

They can be derived by first expressing the localized annihilation operator as
a function of the field $\Psi$ and its conjugated moment $\dot\Psi$. For the 
localized problem we have:
\begin{equation}\label{locfield}
 \Psi(x) = \sum_{{\bf k}} \frac{1}{\sqrt{2\varepsilon_{{\bf k}}}} a_{{\bf k}}
 u_{{\bf k}}({{\bf x}}) \e^{-i \varepsilon_{{\bf k}} t} + {\rm c. c.} 
\end{equation}
where ${\bf k}$ is a vector of three numbers labelling the modes in the region
$V$, $\varepsilon_{{\bf k}}$ is the associated energy and $u_{{\bf k}}$ is 
a complete set of orthogonal wavefunctions over the region $V$:
\begin{eqnarray}
 && \sum_{{\bf k}} u^*_{{\bf k}}({{\bf x}}) u_{{\bf k}}({{\bf x}'}) = 
     \delta^3({{\bf x}}-{{\bf x}'}) \nonumber \\
 && \int_V \d^3 \x \, u^*_{{\bf k}}({{\bf x}}) u_{{\bf k}'}({{\bf x}}) = 
  \delta_{{\bf k},{\bf k}'}    
\end{eqnarray}
and vanishing out of $V$. Inverting the (\ref{locfield}) we have:
\begin{equation}\label{locfield2}
 a_{{\bf k}} =  \frac{i}{\sqrt{2\varepsilon_{{\bf k}}}} \int_V \d^3 \x \,
  u^*_{{\bf k}}({{\bf x}}) \e^{i \varepsilon_{{\bf k}} t } 
  \frac{\stackrel{\leftrightarrow}{\partial}}{\partial t} \Psi(x)
\end{equation}
which are valid at any time. We now enforce the mapping (\ref{mapping}) and 
replace the localized field operators at $t=0$, i.e. in the Schr\"odinger 
representation, with those in the full Hilbert space. In other words, we replace
in (\ref{locfield2}):
\begin{eqnarray}
&& \Psi({\bf x}) \rightarrow \frac{1}{(2\pi)^{3/2}} \int \d^3 \p \,
 \frac{1}{\sqrt{2\varepsilon_{{\bf p}}}} \e^{i {\bf p} \cdot {\bf x}} a_{\bf p} 
 + {\rm c. c.} \nonumber \\
&& \dot\Psi({\bf x}) \rightarrow \frac{1}{(2\pi)^{3/2}} \int \d^3 \p \,
  \frac{-\i\sqrt{\varepsilon_{{\bf p}}}}{\sqrt 2} \e^{i {\bf p} \cdot {\bf x}} 
  a_{\bf p} + {\rm c. c.}
\end{eqnarray}
where $\varepsilon_{{\bf p}} = \sqrt{\p^2+m^2}$, obtaining:
\begin{equation}\label{bogoapp}
 a_{{\bf k}} = \int \d^3 \p \, F({\bf k},{\bf p}) 
 \frac{\varepsilon_{{\bf k}} + \varepsilon_{{\bf p}}}
 {2\sqrt{\varepsilon_{{\bf k}}\varepsilon_{{\bf p}}}}
 \, a_{{\bf p}} + F({\bf k},-{\bf p}) 
 \frac{\varepsilon_{{\bf k}} - \varepsilon_{{\bf p}}}
 {2\sqrt{\varepsilon_{{\bf k}}\varepsilon_{{\bf p}}}}
 \, a_{{\bf p}}^\dagger
\end{equation}
where:
\begin{equation}
 F({\bf k},{\bf p}) = \frac{1}{(2\pi)^{3/2}} 
  \int_V \d^3 \x \, u_{\bf k}^*({\bf x}) \e^{i {\bf p}\cdot{\bf x}}   
\end{equation}
that is the Bogoliubov relations (\ref{bogo}). We observe that the localized 
annihilation operator is a non-trivial combination of annihilation and creation 
operators in the whole space. However, the term in (\ref{bogoapp}) depending
on the creation operator $a^\dagger_{\bf p}$ vanishes in the thermodynamic limit,
as expected. This is most easily shown for a rectangular region, where one has 
$u_{{\bf k}}({{\bf x}}) = \exp[\i {\bf k} \cdot{\bf x}]/\sqrt{V}$;  
${\bf k}$ is given by the eq.~(\ref{complset}) and $\varepsilon_{{\bf k}} = 
\sqrt{{\bf k}^2 + m^2}$. Hence:
\begin{equation}
 \lim_{V \to \infty} F({\bf k},-{\bf p}) \propto 
 \lim_{V \to \infty} \int_V \d^3 \x \, \e^{-i ({\bf p}+{\bf k})\cdot{\bf x}}  
 \propto \delta^3({\bf p}+{\bf k})
\end{equation}
and, consequently $\varepsilon_{{\bf k}} - \varepsilon_{{\bf p}} \to 0$,
which make the second term in (\ref{bogoapp}) vanishing. 

We can use the eq.~(\ref{bogoapp}) to work out linear relations between Fock space
states of the localized problem and asymptotic Fock space states. These can be 
obtained enforcing the destruction of the localized vacuum state:
$$
 a_{\bf k} \ket{0}_V = 0  
$$
and writing:
\begin{equation}
 \ket{0}_V = \alpha_0 \ket{0} + \int \d^3 \rm p \, \alpha_1({\bf p}) \ket{{\bf p}} +
 \int \d^3 \p_1 \d^3 \p_2 \, \alpha_2({\bf p}_1,{\bf p}_2) \ket{{\bf p}_1,{\bf p}_2} + \ldots
\end{equation}
Working out such relations is beyond the scope of this paper.

\section{Multiparticle state with $N$ particles}

In order to prove eq.~(\ref{nparticles4}) for neutral particles, the first step
is to realize that a general expansion of:
\begin{equation}\label{starting}
  \bra{0} a_1 \ldots a_N \Pro_V  a_N^\dagger \ldots a_1^\dagger \ket{0}
\end{equation}
($a_i$ is a shorthand for $a(p_i)$) must yield, on the basis of what shown for 
the case of 2 particles, a sum of terms like these:
\begin{equation}\label{genterm}
  \int_V \d^3 \x_1 \ldots \int_V \d^3 \x_N 
  \prod_{n,m=1}^N \e^{\pm i {\bf p}_n \cdot {\bf x}_m}
  f(\varepsilon_1,\ldots,\varepsilon_N) \bra{0}\Pro_V\ket{0}
\end{equation}
where $f$ is a generic function involving a sum of ratios or product of any 
number of particle energies and $\pm$ stands for either a $+$ or a $-$ sign.
That (\ref{genterm}) ought to be the final expression can be envisaged on the 
basis of a repeated application of the equations (\ref{scfield4}),(\ref{scfield19}),
(\ref{scfield14a}),(\ref{scfield13a}) in turn.
The eq.~(\ref{genterm}) can also be written as:
\begin{equation}\label{genterm2}
  \int_V \d^3 \x_1 \ldots \int_V \d^3 x_N \e^{\i \sum'_n {\bf p}_n \cdot {\bf x}_1}
 \ldots \e^{\i \sum'_n {\bf p}_n \cdot {\bf x}_n}
  f(\varepsilon_1,\ldots,\varepsilon_N) \bra{0}\Pro_V\ket{0}
\end{equation}
where $\sum'$ stands for an algebraic sum with terms having either sign.

If we now take the thermodynamic limit $V \to \infty$ of (\ref{starting}), one
has $\Pro_V \to {\sf I}$, thence:
\begin{eqnarray}\label{tlimit}
&& \lim_{V \to \infty} 
\bra{0} a_1 \ldots a_N \Pro_V  a_N^\dagger \ldots a_1^\dagger \ket{0} =
\bra{0} a_1 \ldots a_N a_N^\dagger \ldots a_1^\dagger \ket{0} \nonumber \\
&=& \sum_{\rho \in S_N} \prod_{n=1}^N \delta^3({\bf p}_n - {\bf p}_{\rho(n)})
\end{eqnarray}
This tells us that the function $f$'s in each term (\ref{genterm}) must reduce
to a trivial factor $1$ because they would otherwise survive in the thermodynamic
limit, being a factor depending only on the $p_n$'s. Moreover, since the 
thermodynamic limit involves only Dirac's $\delta$'s with differences of two momenta 
as argument, there can be only difference of two momenta as argument of exponential 
functions in (\ref{genterm2}). Finally, by comparing the eq.~(\ref{genterm2}) 
with the (\ref{tlimit}), we conclude that the only possible expression at finite $V$
is: 
\begin{equation}\label{final}
\bra{0} a_1 \ldots a_N \Pro_V  a_N^\dagger \ldots a_1^\dagger \ket{0}=
\bra{0} \Pro_V \ket{0} \sum_{\rho \in S_N} \prod_{n=1}^N \frac{1}{(2\pi)^3} 
 \int_V \d^3 \x_n \, \e^{-\i ({\bf p}_n - {\bf p}_{\rho(n)})\cdot {\bf x}_n}
\end{equation}
which is precisely (\ref{nparticles4}).

For a state with $N_+$ particles and $N_-$ antiparticles, the validity of 
eq.~(\ref{twospecies}) can be argued with a similar argument, i.e. by constraining
the form of general terms like (\ref{genterm}) taking advantage of the limit
$V \to \infty$. In the case of particles and antiparticles, the thermodynamic
limit tells us that:
\begin{eqnarray}\label{tlimit2}
&& \lim_{V \to \infty} 
\bra{0} a_1 \ldots a_{N_+} b_1 \ldots b_{N_-}\Pro_V  a_{N_+}^\dagger 
 \ldots a_1^\dagger b_{N_-}^\dagger \ldots b_1^\dagger \ket{0} \nonumber \\
&=& \bra{0} a_1 \ldots a_{N_+} b_1 \ldots b_{N_-} a_{N_+}^\dagger 
 \ldots a_1^\dagger b_{N_-}^\dagger \ldots b_1^\dagger \ket{0}
  \nonumber \\
&=& \sum_{\rho_+ \in S_{N_+}} \prod_{n_+=1}^{N_+} \delta^3({\bf p}_{n_+} 
 - {\bf p}_{\rho_+(n_+)})
 \sum_{\rho_- \in S_{N_-}} \prod_{n_-=1}^{N_-} \delta^3({\bf p}_{n_-} 
 - {\bf p}_{\rho_-(n_-)})
\end{eqnarray}
and this determines the form of the expression:
$$
\bra{0} a_1 \ldots a_{N_+} b_1 \ldots b_{N_-}\Pro_V  a_{N_+}^\dagger 
 \ldots a_1^\dagger b_{N_-}^\dagger \ldots b_1^\dagger \ket{0}
$$
at finite $V$ to be (\ref{twospecies}). The absence of integrals mixing particles 
with antiparticles momenta such as:
$$
  \int \d^3 \x_i \e^{\i ({\bf p}_{n_+} - {\bf p}_{n_-}) \cdot {\bf x}_i}
$$
is owing to the absence of such differences as arguments of $\delta$'s
in eq.~(\ref{tlimit2}).


\end{document}